\documentclass[12pt]{article}
\usepackage{epsfig}
\usepackage[dvips]{color}
\begin{document}
\renewcommand{\theequation}{\arabic{section}.\arabic{equation}}

\def\be{\begin{equation}}
\def\ee{\end{equation}}
\def\bq{\begin{equation}}
\def\eq{\end{equation}}
\def\bqa{\begin{eqnarray}}
\def\eqa{\end{eqnarray}}
\def\roughly#1{\mathrel{\raise.3ex
\hbox{$#1$\kern-.75em\lower1ex\hbox{$\sim$}}}}
\def\lsim{\roughly<}
\def\gsim{\roughly>}
\def\llgm{\left\lgroup\matrix}
\def\rrgm{\right\rgroup}
\def\vectrl #1{\buildrel\leftrightarrow \over #1}
\def\partrl{\vectrl{\partial}}
\def\gslash#1{\slash\hspace*{-0.20cm}#1}
%\renewcommand{\theequation}{\arabic{section}.\arabic{equation}}

%\begin{flushright}
%November 19, 2012
%\end{flushright}

%{\bf Preliminary version\hfill February 8, 2013}
%\vspace*{2cm}

\begin{center}
%{\bf Version 3 (27.01.2014)\\
%{\bf Version  (May 22, 2016)\\
%{\color{red}{\bf ``Small corrections 2'' }}
{\bf
Color Dipole Picture of Deep Inelastic Scattering,\\
Revisited.}%\footnote{Supported by Deutsche 
%Forschungsgemeinschaft, contract number schi 189/6-2}\footnote{email:
%Dieter.Schildknecht@physik.uni-bielefeld.de}
\end{center}
 \vspace {0.5 cm}
\begin{center}
{\bf  Masaaki Kuroda}\\[2.5mm]
Center for Liberal Arts, Meijigakuin University\\ [1.2mm] 
Yokohama, Japan\\ [3mm] 
{\bf Dieter Schildknecht} \\[2.5mm]
Fakult\"{a}t f\"{u}r Physik, Universit\"{a}t Bielefeld \\[1.2mm] 
D-33501 Bielefeld, Germany \\[1.2mm]
and \\[1.2mm]
Max-Planck Institut f\"ur Physik (Werner-Heisenberg-Institut),\\[1.2mm]
F\"ohringer Ring 6, D-80805, M\"unchen, Germany
\end{center}

\vspace{2 cm}

\baselineskip 20pt

\begin{center}
{\bf Abstract}
\end{center}
Based upon the color-dipole picture, we provide closed analytic
expressions for the longitudinal and the transverse photoabsorption
cross sections at low values of the Bjorken variable of $x \lsim 0.1$.
We compare with the experimental data for the longitudinal-to-transverse
ratio of the (virtual) photoabsorption cross section and with
our previous fit to the experimental data for the total photoabsorption
cross section. Scaling in terms of the low-x scaling variable
$\eta (W^2, Q^2)$ is analyzed in terms of the reduced cross section
of deep inelastic scattering.

\vfill\eject

\section{Introduction}
\renewcommand{\theequation}{\arabic{section}.\arabic{equation}}
\setcounter{equation}{0}

The process of deep inelastic scattering (DIS) at low values of the Bjorken
variable, $x \cong Q^2/W^2 \lsim 0.1$, where $Q^2 \ge 0$ and $W^2$ refer
to the virtuality of the photon and the photon-nucleon center of mass
energy squared, is determined by the virtual dissociation of the photon 
into hadronic
vector states that subsequently interact with the nucleon (generalized
vector dominance (GVD)) \cite{Sakurai}, \cite{FR}, \cite{APP}. 
In QCD, the hadronic vector states are quark-antiquark
states that interact as color dipoles with the gluon field in the nucleon
by exchange of (at least) two gluons that form a color-neutral 
state (color dipole picture (CDP))
\cite{Low}, \cite{Nikolaev}, \cite{Cvetic}, \cite{DIFF2000},
\cite{1108}.
The color-gauge invariant interaction with the gluon field in the
nucleon -- without specific parameterization of this interaction
\cite{PRD85}  -- implies 
color transparency and saturation,\footnote{Compare also the recent
reviews in ref. \cite{Lanz1}} respectively dependent on
the relative magnitude of $Q^2$ and $W^2$ within the region of 
$x \lsim 0.1$.

The quantitative analysis of the experimental data on the photon absorption
cross section, or, equivalently, the proton structure functions, requires
a fit \cite{DIFF2000}, \cite{1108} \cite{PRD85}
to the experimental data on DIS
based on the small number of two to four
free parameters, 
the number of parameters dependent on which ones are considered to
be fixed by theoretical considerations.

In Section 2, we present concise and simple closed analytic expressions
for the longitudinal and transverse photoabsorption cross sections. In
Section 3, our results are summarized in a convenient form to be used 
in elaborate fits by experimentalists to the body of experimental data. 
We compare our theoretical predictions for the total photoabsorption 
cross section and the longitudinal-to-transverse ratio with experimental
data. Some conclusions will be drawn in Section 4. Technical details
are given in Appendices A to C.

\section{The Photoabsorption Cross Section in the Color Dipole Picture,
Theory}
\renewcommand{\theequation}{\arabic{section}.\arabic{equation}}
\setcounter{equation}{0}

In Section 2.1, we summarize the essential results from the color dipole
picture (CDP). In Section 2.2 various refinements will be presented.

\subsection{The color dipole picture, formulation from 2000.}
The approach of the color dipole picture (CDP) to deep inelastic 
scattering\footnote{Compare ref. \cite{KS_EPJC37} for the
application of the 
CDP to deeply virtual Compton scattering and vector meson production,
and ref. \cite{KS_PRD88} for the treatment of ultra-high-energy
neutrino-nucleon scattering.}
at low $x \cong Q^2/W^2 \lsim 0.1$ may be summarized by the 
photoabsorption cross section \cite{Cvetic}, \cite{Nikolaev},
\be
\sigma_{\gamma^*_{L,T}} (W^2,Q^2) = \int dz \int d^2 \vec r_\bot \bigg|
\psi_{L,T} (\vec r_\bot, z (1-z),Q^2) \bigg|^2 \sigma_{(q \bar q)p}
(\vec r_\bot, z(1-z), W^2),
\label{2.1}
\ee
supplemented by the representation of the dipole cross section,
\be
\sigma_{(q \bar q)p} (\vec r_\bot, z(1-z), W^2) = \int d^2 \vec l_\bot
\tilde{\sigma} (\vec l_\bot^{~2}, z(1-z),W^2) (1-e^{-i\vec l_\bot \vec r_\bot})
\label{2.2}
\ee
that guarantees the gauge-invariant interaction of the 
quark-antiquark $(q \bar q)$ color
dipole with the gluon field in the proton via two-gluon coupling.
The ``photon wave function'' squared, $\vert \psi_{L,T} (\vec r_\bot, z(1-z),
Q^2) \vert^2$, in (\ref{2.1}) is determined by QED. It gives the probability
for the fluctuation of the (virtual) photon of virtuality $Q^2 \ge 0$ into
a $q \bar q$ dipole state of configuration $(\vec r_\bot, z (1-z))$. The
variable $\vec r_\bot$ in (\ref{2.1}) and (\ref{2.2}) determines the transverse
$q \bar q$-separation variable of the $q \bar q$ state, and $0 \le z \le 1$ 
characterizes the
longitudinal momentum partition between quark and antiquark in that state.

The right-hand side in (\ref{2.1}) contains the (required) factorization
of the $Q^2$-dependent photon wave function and the $W$-dependent dipole
cross section: the photon of virtuality $q^2 = - Q^2 \le 0$ virtually
dissociates (GVD \cite{Sakurai}, \cite{FR}, \cite{APP}), 
or fluctuates in modern jargon, into $q \bar q$ states of 
masses\footnote{The appropriate Fourier transform introduces the
transverse $q \bar q$-separation variable $\vec r_\bot$ and the 
longitudinal momentum partition, $z$.}
$M_{q \bar q} > 0$
that propagate and interact with the proton at the center-of-mass energy $W$.
In ref. \cite{Cvetic} the formulation of (\ref{2.1}) with (\ref{2.2})
of the CDP was explicitly derived from GVD supplemented by the QCD-based
color-dipole structure of the interacting $q \bar q$ vector states.The
dependence on $W$ of the dipole cross section in (\ref{2.1}) is a strict
consequence from the mass dispersion-relation (compare Appendix B)
of GVD; the necessary dependence
on $W$ in (\ref{2.1})
(rather than a dependence on $x \cong Q^2/W^2$ \cite{Nikolaev}), using 
functional methods of quantum field theory, was more recently elaborated
upon in great detail in ref. \cite{Ewerz}.

The representation of the photoabsorption cross section (\ref{2.1}), in
conjunction with the color-gauge-invariant representation of the
$q \bar q$-dipole-proton interaction (\ref{2.2}), implies the essential
qualitative feature of the experimental results: the scaling of the total
photoabsorption cross section $\sigma_{\gamma^*p} (W^2,Q^2) = 
\sigma_{\gamma^*p} (\eta (W^2,Q^2))$ in the low-x scaling variable
$\eta (W^2, Q^2)$ \cite{DIFF2000}, \cite{1108}, \cite{PRD85}, 
see (\ref{2.5}) below, as $1/\eta (W^2,Q^2)$ for
$\eta (W^2,Q^2) \gg 1$, (``color transparency'') and as $\ln (1/\eta (W^2,Q^2))$
for $\eta (W^2,Q^2) \ll 1$ (``saturation''). As elaborated upon in detail in
ref. \cite{PRD85}, no  parameter-dependent explicit  ansatz for the
dipole cross section (\ref{2.2}) is required to arrive at this general
conclusion. 

For an explicit quantitative description of the experimental results on the
total photoabsorption cross sections for longitudinally and transversely
polarized photons, a parameter-dependent specification of the dipole cross
section is required.
In ref. \cite{DIFF2000}, the general expression for the photoabsorption cross
section in (\ref{2.1}) with (\ref{2.2}) is supplemented by the ansatz
\be
\tilde{\sigma} \left( \vec l_\bot^{~2}, z(1-z), W^2 \right) =
\frac{\sigma^{(\infty)} (W^2)}{\pi} \delta \left( \vec l_\bot^{~2} -z (1-z)
\Lambda^2_{sat} (W^2) \right)
\label{2.3}
\ee
that implies
\be
\sigma_{(q \bar q)p} (\vec r_\bot, z(1-z), W^2) = \sigma^{(\infty)} (W^2)
\left(1 - J_0 (r_\bot \sqrt{z(1-z)} \Lambda_{sat} (W^2) \right).
\label{2.4}
\ee
In (\ref{2.3}) and (\ref{2.4}), $\sigma^{(\infty)} (W^2)$ denotes a weakly 
(logarithmically) on $W^2$
dependent cross section of hadronic magnitude to be elaborated upon later,
$\vec l_\bot$ denotes the transverse part of the three-momentum of the 
gluon absorbed by the $q \bar q$ dipole state, 
and $\Lambda^2_{sat} (W^2)$ denotes 
the ``saturation scale'' that increases with a small power of $W^2$.
The ansatz (\ref{2.3}) implies, and is motivated by, a required increase
of the effective transverse momentum of the absorbed gluon with increasing 
energy $W$. The function $J_0 \left( r_\bot 
\sqrt{z(1-z)} \Lambda_{sat} (W^2) \right)$ in (\ref{2.4})
is the zero-order Bessel function.

We also note the low-x scaling variable, $\eta (W^2,Q^2)$, given by
\cite{DIFF2000}
\be
\eta (W^2,Q^2) = \frac{Q^2 + m^2_0}{\Lambda^2_{sat} (W^2)}.
\label{2.5}
\ee
In addition to (\ref{2.5}), it is useful to introduce the ratio
\be
\mu (W^2) = \frac{m^2_0}{\Lambda^2_{sat} (W^2)}.
\label{2.6}
\ee
The low-x scaling variable $\eta (W^2,Q^2)$ becomes $\eta (W^2,Q^2) =
Q^2/\Lambda^2_{sat} (W^2) + \mu (W^2) \ge \mu (W^2)$.
The mass $m_0$ in (\ref{2.5}) and (\ref{2.6})
denotes the effective onset of hadron production in
$e^+e^-$ annihilation to hadrons, and, from quark-hadron duality \cite{DS-FS}, 
we have $0 < m^2_0 < m^2_\rho$, where $m_\rho$ denotes the 
$\rho$-meson mass.\footnote{For heavy flavors, like charm, the mass scale
$m^2_0$ must be appropriately modified.}

The evaluation of the photoabsorption cross section (\ref{2.1}), upon
insertion of (\ref{2.4}), leads to \cite{DIFF2000}
\be
\sigma_{\gamma^*_{L,T}p} (W^2,Q^2) = \frac{\alpha R_{e^+e^-}}{3 \pi}
\sigma^{(\infty)} (W^2) I_{L,T} \left( \eta (W^2,Q^2), \mu (W^2) \right).
\label{2.7}
\ee
In (\ref{2.7}), $R_{e^+e^-} = 3 \sum_q Q^2_q$, where the sum runs over
the actively contributing quark flavors, and $Q_q$ denotes the quark charge.

It turns out that the relevant region of $\mu (W^2)$ fulfills the bound
$\mu (W^2) < 1$. Under this assumption $I_{L,T} (\eta (W^2,Q^2), \mu
(W^2))$ becomes \cite{DIFF2000}
\be
I_{L,T} (\eta (W^2,Q^2),\mu (W^2)) = I^{(1)}_{L,T} (\eta (W^2,Q^2), \mu (W^2))
\left( 1+0 \left( \mu (W^2)\right) \right),
\label{2.8}
\ee
where $I^{(1)}_L (\eta (W^2,Q^2), \mu (W^2))$ and $I^{(1)}_T 
(\eta (W^2,Q^2), \mu (W^2))$
are given by
\bqa
&& I^{(1)}_L (\eta, \mu) = \frac{\eta - \mu}{\eta} \nonumber \\
&& \times \left( 1 - 
\frac{\eta}{\sqrt{1+4 (\eta - \mu)}} \ln 
\frac{\eta (1+ \sqrt{1+4(\eta - \mu)})}{4 \mu-1-3\eta + 
\sqrt{(1+4(\eta - \mu))((1+\eta)^2 - 4 \mu)}} \right), \nonumber \\
&& I^{(1)}_T (\eta, \mu)  =  \frac{1}{2} \ln \frac{\eta-1 + 
\sqrt{(1+\eta)^2 - 4 \mu}}{2 \eta} 
- \frac{\eta - \mu}{\eta} + \frac{1+2(\eta - \mu)}
{2 \sqrt{1+4 (\eta - \mu)}}  \nonumber \\
&&~~~~~~~~~~~~~ \times  \ln \frac{\eta (1 + \sqrt{1+4 (\eta - \mu)})}
{4 \mu -1-3\eta + \sqrt{(1+4(\eta-\mu))((1+\eta)^2-4 \mu)}}.
\label{2.9}
\eqa

We note the photoproduction $(Q^2 = 0)$ limit of (\ref{2.7}) with
(\ref{2.9}). Inserting $\eta = c \mu (W^2)$ into (\ref{2.9}), where
$c = const \ge 1$ and $0 < \mu (W^2) < 1$, a careful evaluation of the
photoproduction limit of $c \to 1$ leads to 
\bqa
&& \lim_{\eta \to \mu}  \sigma_{\gamma^*_L p} (W^2,Q^2) = 0,\nonumber \\
\sigma_{\gamma p} (W^2) \equiv && \lim_{\eta \to \mu} \sigma_{\gamma^*_Tp}
(W^2,Q^2) 
= \frac{\alpha R_{e^+e^-}}{3 \pi} \sigma^{(\infty)} (W^2) \ln
\frac{1}{\mu}.
\label{2.10}
\eqa

In the limit of very high energy, $\mu (W^2) \ll 1$, (\ref{2.9}) 
may be further simplified.
We note that $\mu (W^2) \le \eta (W^2,Q^2) \le \eta_{Max} (W^2)$, 
where $\eta_{Max} (W^2)$ is
determined by the required restriction to low values 
of $x \le x_0 \lsim 0.1$, or
$Q^2 \le x_0W^2$, i.e.
\be
\eta(W^2,Q^2) \le \eta_{Max} (W^2) = \frac{x_0W^2}{\Lambda^2_{sat} (W^2)}.
\label{2.11}
\ee
With $\mu (W^2) \ll 1$ and $\eta (W^2,Q^2) \ge \mu (W^2)$, 
upon making use of the identity
\bqa
2 \ln \frac{\sqrt{1+4 \eta} +1}{\sqrt{1+4 \eta} -1} & = & \ln
\frac{(1+\eta) \sqrt{1+4 \eta} + 1 + 3 \eta}{\eta (\sqrt{1 + 4 \eta} -1)} =
\nonumber \\
& = & \ln \frac{\eta (1 + \sqrt{1 + 4 \eta})}{(1 + \eta) \sqrt{1 + 4\eta}
-1 - 3 \eta},
\label{2.12}
\eqa
and of the definition
\be
I_0 (\eta) = \frac{1}{\sqrt{1 + 4 \eta}} \ln 
\frac{\sqrt{1 + 4 \eta} +1}{\sqrt{1 + 4 \eta} -1},
\label{2.13}
\ee
we find that (\ref{2.9}) becomes
\bqa
I^{(1)}_L (\eta, \mu) & = & \frac{\eta - \mu}{\eta} (1 - 2 \eta I_0 (\eta)),
\nonumber \\
I^{(1)}_T (\eta, \mu) & = & I_0 (\eta) - \frac{\eta - \mu}{\eta}
(1 - 2 \eta I_0(\eta)),
\label{2.14}
\eqa
and, accordingly,
\be
I^{(1)}_L (\eta, \mu) + I^{(1)}_T (\eta, \mu) = I_0 (\eta).
\label{2.15}
\ee

With (\ref{2.14}), (\ref{2.15})  and (\ref{2.8}), the
longitudinal and the transverse parts of the 
photoabsorption cross section (\ref{2.7}), and
the total cross section
\be
\sigma_{\gamma^*p} (W^2,Q^2) = \sigma_{\gamma^*_L p} (W^2,Q^2) +
\sigma_{\gamma^*_T p} (W^2,Q^2),
\label{2.16}
\ee
explicitly become,
\bqa
\sigma_{\gamma^*_L p} (W^2,Q^2) & = & \frac{\alpha R_{e^+e^-}}{3 \pi}
\sigma^{(\infty)} (W^2) \frac{\eta - \mu}{\eta} (1- 2 \eta I_0 (\eta)),
\nonumber\\
\sigma_{\gamma^*_T p} (W^2,Q^2) & = & \frac{\alpha R_{e^+e^-}}{3 \pi}
\sigma^{(\infty)} (W^2) \left( I_0 (\eta) - \frac{\eta - \mu}{\eta}
\left( 1- 2 \eta I_0 (\eta)\right) \right),
\nonumber \\
\sigma_{\gamma^* p} (W^2,Q^2) & = & \frac{\alpha R_{e^+e^-}}{3 \pi}
\sigma^{(\infty)} (W^2) I_0 (\eta).
\label{2.17}
\eqa
Under the approximation of $\sigma^{(\infty)} (W^2) \cong ~const$
(actually $\sigma^{(\infty)} (W^2)$ varies as $\ln W^2$, see below),
we have low-x scaling behavior \cite{DIFF2000},
\be
\sigma_{\gamma^*p} (W^2,Q^2) = \sigma_{\gamma^*p} \left( \eta(W^2,Q^2)\right).
\label{2.18}
\ee
The longitudinal cross section in (\ref{2.17})
vanishes in the $Q^2 = 0$ photoproduction limit 
of $\eta (W^2,Q^2) = \mu (W^2)$. For $\eta \gg \mu$, there is low-x
scaling separately for the longitudinal and transverse parts of the
cross section, $\sigma_{\gamma^*_Lp}
(W^2,Q^2) = \sigma_{\gamma^*_Lp} \left( \eta (W^2,Q^2) \right)$ and
$\sigma_{\gamma^*_T} (W^2,Q^2) = \sigma_{\gamma^*_T} \left( \eta (W^2,Q^2)
\right)$.

It will be illuminating to examine the result (\ref{2.17}) for different 
limits of $\eta$ and $\mu$.
\begin{itemize}
\item[i)] The limit of $\mu \ll 1$ combined with $\eta \gg 1$ (and $\eta
\le \eta_{Max}$). This limit corresponds to high energy $W$ and relatively
large photon virtualities, $Q^2 \gg \Lambda^2_{sat} (W^2)$. Employing the
expansion,
\bqa
I_0 (\eta) & = & \frac{2}{1 + 4 \eta} + \frac{2}{3}
\frac{1}{(1 + 4 \eta)^2} + \frac{2}{5} \frac{1}{(1 + 4 \eta)^3} + ...,
~~~~~~(\eta \gg 1) \nonumber \\
& = & \frac{1}{2 \eta} \left( 1 - \frac{1}{6 \eta} - \frac{11}{120 \eta^2}
\right)
+ 0 \left( \frac{1}{\eta^4} \right)
\label{2.19}
\eqa
from (\ref{2.14}) and (\ref{2.15}) we find
\be
I_L^{(1)} (\eta \gg 1, \mu \ll 1) = \frac{1}{6 \eta} + 0 
\left( \frac{1}{\eta^2} \right),
\label{2.20}
\ee
and
\be
I_T^{(1)} (\eta \gg 1, \mu \ll 1) = \frac{1}{3 \eta} + 0 
\left( \frac{1}{\eta^2} \right),
\label{2.21}
\ee
as well as
\be
I_L^{(1)} (\eta \gg 1, \mu \ll 1) + I_T^{(1)} (\eta \gg 1, \mu \ll 1) =
\frac{1}{2 \eta} + 0 \left( \frac{1}{\eta^2} \right).
\label{2.22}
\ee
In this limit of $\eta \gg 1$, the ratio of the longitudinal to the
transverse photoabsorption cross section becomes $1/2$, \cite{DIFF2000},
\cite{PRD85},
\be
R = \frac{\sigma_{\gamma^*_L p} \left( \eta (W^2,Q^2)
  \right)}{\sigma_{\gamma^*_T p} \left( \eta (W^2,Q^2) \right)} = \frac{1}{2}.
\label{2.23}
\ee
The result (\ref{2.23}) is a direct consequence of the explicit form of
the photon wave function in (\ref{2.1}). A dependence of the dipole
cross section on the product $r_\bot \sqrt{z(1-z)}$ implies \cite{DIFF2000},
\cite{PRD85} helicity independence, the equality of the dipole cross sections
(in the limit of $r^2_\bot \to 0$) for transversely polarized 
$q \bar q$ states
(originating from $\gamma^*_T$) and longitudinally polarized ones 
(originating from 
$\gamma^*_L$). The enhanced transverse size of transverse relative to
longitudinal $q \bar q$ states calls for a refinement of the simplifying
assumption of helicity independence contained in the ansatz (\ref{2.3})
and (\ref{2.4}). The ratio $R$ in (\ref{2.23}) then becomes \cite{Ku-Schi},
\cite{PRD85}
\be
R = \frac{1}{2 \rho},
\label{2.24}
\ee
with $\rho = const > 1$. Compare Section 2.2 and Appendix A.
\item[ii)] The limit of $\eta = c \mu$ with $c = const \ge 1$, and
\begin{enumerate}
\item[a)] $\mu \ll 1$ fixed, $c \to 1$, the limit of
$Q^2 = 0$ photoproduction at fixed energy $W$,
\item[b)] $\mu \to 0,~\eta = c \mu \to 0,~c > 1$ fixed, the limit of 
$W^2 \to \infty$ with $Q^2 = (c-1) m^2_0 > 0$ fixed.
\end{enumerate}
Evaluating $I_0 (\eta)$ in (\ref{2.13}) in the
limit of $\eta = c \mu \ll 1$, we find the leading term of
\be
I_0 (\eta) \cong \left( \ln \frac{1}{\eta} \right) 
\left( 1-2\eta \left(1-\frac{1}{\ln \frac{1}{\eta}}\right) \right) 
\cong \ln \frac{1}{c \mu},
~~~~~(\eta = c \mu \ll 1).
\label{2.25}
\ee
Substitution into (\ref{2.14}) and (\ref{2.15}) yields
%\newpage
\bqa
I_L^{(1)} (\eta = c \mu, c\mu \ll 1) && = \frac{c-1}{c} - 2 \mu (c-1)
\ln \frac{1}{c \mu} \nonumber \\
&& = \left\{ \begin{array}{l@{\quad\quad}l}
0 & {\rm for}~~c=1, (Q^2 =0, ~ {\rm photopr.}),\\
  & ~ \\
{\displaystyle \frac{c-1}{c}} 
& {\rm for}~~c>1~{\rm fixed}~, ~\eta = c \mu \to 0\\
&(W^2 \to \infty,~Q^2 > 0~{\rm fixed}),
\end{array} \right.
\label{2.26}
\eqa
and
\be
I_T^{(1)} (\eta = c \mu,~c\mu \ll 1) = 
\left\{ \begin{array}{l@{\quad\quad}l}
{\displaystyle \ln \frac{1}{\mu}} & {\rm for}~~c=1, (Q^2 = 0,~{\rm photopr.}),\\
  & ~ \\
{\displaystyle \ln \frac{1}{c \mu} - \frac{c-1}{c}}
& {\rm for}~~c>1~{\rm fixed},~
\eta = c \mu \to 0\\
& (W^2 \to \infty,~Q^2 > 0~{\rm fixed}).
\end{array} \right.
\label{2.27}
\ee
\end{itemize}
The $Q^2 = 0$, photoproduction, results in (\ref{2.26}) and (\ref{2.27}), 
based on the very-high-energy $(\mu (W^2) \ll 1)$ representation (\ref{2.17}),
agree with the result (\ref{2.10}) obtained from (\ref{2.9}) with
$0 < \mu (W^2) < 1$.

 In the limit of $W^2 \to \infty$, but $Q^2 > 0$ 
fixed i.e. for $c > 1$ fixed and $\eta = c \mu \to 0$, upon combining
(\ref{2.26}) and (\ref{2.27}), we find \cite{DIFF2000, 1108, PRD85}
\footnote{Equivalently, one finds (\ref{2.28}) from (\ref{2.9}) by inserting
$\mu = c \mu$ into (\ref{2.9}), and evaluating the limit of $\mu \to 0$ with
$c > 1$ fixed.}
\be
\lim_{W^2 \to \infty \atop{Q^2 > 0~{\rm fixed}}} \frac{\sigma_{\gamma^*p}
(W^2,Q^2)}{\sigma_{\gamma p} (W^2)} = \lim_{W^2 \to \infty \atop {Q^2 > 0~
{\rm fixed}}} \frac{\sigma_{\gamma^*_T p} (W^2,Q^2)}{\sigma_{\gamma p} (W^2)} =
\lim_{\mu \to 0} \frac{\ln \frac{1}{c \mu}}
{\ln \frac{1}{\mu}} = 1.
\label{2.28}
\ee
At any $Q^2 > 0$, if the energy $W$ is sufficiently large (``saturation
limit''), 
the ratio of the photoabsorption cross section 
at finite $Q^2$ over the
$Q^2 = 0$ photoabsorption cross section, $\sigma_{\gamma p} (W^2)$, becomes
equal to unity. Only the transverse part of the photoabsorption cross
section contributes to the ratio (2.28) in this limit of $W^2 \to \infty$ at
$Q^2 > 0$ fixed.

\subsection{Refinements}
The formulation of the CDP for the photoabsorption cross section in
(\ref{2.1}) to (\ref{2.4}) implicitly contains contributions from
$q \bar q$ fluctuations of the (virtual) photon of unlimited mass,
$M_{q \bar q} \to \infty$. The life time $\tau$ of a $q \bar q$ fluctuation,
in the rest frame of the target proton of mass $M_p$ \cite{Nieh}, 
\cite{Bilchak}, \cite{Ewerz}
\be
\tau \cong \frac{1}{\Delta E} = \frac{1}{x + \frac{M^2_{q \bar q}}{W^2}}
\frac{1}{M_p},
\label{2.29}
\ee
with increasing mass $M_{q \bar q}$ decreases strongly, however,
at any fixed energy $W$ and any fixed value of $x \cong Q^2/W^2 \lsim 0.1$.
To assure the sufficiently long lifetime of $\tau \gg 1/M_p$ necessary
for the validity of the CDP of diffractive $(q \bar q) p$ forward scattering,
the masses of the contributing $q \bar q$ fluctuations must be
restricted by a W-dependent upper limit,  
$M^2_{q \bar q} \le m^2_1 (W^2)$.

In ref. \cite{PRD85} we gave a formulation of the photoabsorption cross
section in the CDP that incorporates the required upper 
bound\footnote{An upper bound,$M^2_{q \bar q} \le m^2_1$, was also
previously introduced in ref. \cite{KS_PRD67}.} on 
$M^2_{q \bar q}$,
\be
M^2_{q \bar q} \le m^2_1 (W^2) = \xi \Lambda^2_{sat} (W^2).
\label{2.30}
\ee
Adjustment to the experimental data on DIS showed consistency with the
upper limit
(\ref{2.30}) for a constant value of $\xi$ of magnitude of approximately
\be
\xi = {\rm const} \cong 130.
\label{2.31}
\ee
The modifications of the longitudinal and the transverse photoabsorption
cross sections (\ref{2.7}) and (\ref{2.17}) implied by the constraint 
(\ref{2.30}) can
be cast into simple factors that depend on the ratio
\be
u = u (\eta (W^2,Q^2)) = \frac{\xi}{\eta (W^2,Q^2)}.
\label{2.32}
\ee
The factors, $G_L (u)$ and $G_T (u)$, will be specified
below.

The refined formulation of the CDP in ref. \cite{PRD85} also includes the
transverse-size enhancement \cite{Ku-Schi}
of transversely relative to longitudinally
polarized $q \bar q$ fluctuations by the factor $\rho$ that was mentioned
in (\ref{2.24}).
The factor $\rho$ enters the transverse photoabsorption cross section
in (\ref{2.17}) via the replacement
\be
I^{(1)}_T (\eta, \mu) \to I^{(1)}_T \left( \frac{\eta}{\rho}, \frac{\mu}{\rho}
\right),
\label{2.33}
\ee
compare (\ref{A45}) in Appendix A.

Taking into account the constraint (\ref{2.30}), as well as the
transverse-size enhancement factor $\rho$ according to (\ref{2.33}),
the cross sections in (\ref{2.7}), expressed in terms of the functions
$I^{(1)}_L~(\eta, \mu)$ and $I^{(1)}_T (\eta, \mu)$ in 
(\ref{2.9}) or (\ref{2.14})
become\footnote{The compact form of the photoabsorption cross sections in
  (\ref{2.33a}) is a novel result of the present investigation. Compare
Appendix A for its derivation and for the connection with our previous 
results in ref. \cite{PRD85}}
%\newpage
\bqa
\sigma_{\gamma^*_L p} (W^2,Q^2) & = & \frac{\alpha R_{e^+e^-}}{3 \pi} 
\sigma^{(\infty)} (W^2) I_L^{(1)} (\eta, \mu) G_L (u), \nonumber \\
\sigma_{\gamma^*_T p} (W^2,Q^2) & = & \frac{\alpha R_{e^+e^-}}{3 \pi} 
\sigma^{(\infty)} (W^2) I_T^{(1)} \left(\frac{\eta}{\rho},
\frac{\mu}{\rho}\right) G_T (u).
\label{2.33a}
\eqa
For details we refer to Appendix A, specifically see (\ref{A38}) and
(\ref{A45}).
The functions $G_L (u)$ and $G_T (u)$ are given by \cite{PRD85}
\bqa
G_L (u) & = &\frac{2u^3 + 6u^2}{2 (1+u)^3} \simeq
\left\{ \begin{array}{l@{\quad,\quad}l}
3 u^2 & (u \ll 1),\\
1 - \frac{3}{u^2} & (u \gg 1),
\end{array} \right.\nonumber \\
G_T (u) & = & \frac{2u^3 + 3 u^2 + 3u}{2 (1+u)^3} \simeq
\left\{ \begin{array}{l@{\quad,\quad}l}
\frac{3}{2} u & (u \ll 1), \\
1 - \frac{3}{2u} & (u \gg 1),
\end{array} \right.
\label{2.34}
\eqa
where, according to (\ref{2.32}), the limits of $u(\eta(W^2,Q^2)) \ll 1$ 
and $u(\eta(W^2,Q^2)) \gg 1$
correspond to the large-$\eta$ limit of
$\eta (W^2,Q^2) \gg \xi$ and the small-$\eta$ limit of 
$\eta (W^2,Q^2) \ll \xi$,
respectively. According to (\ref{2.34}), in the large-$\eta$ limit, 
$\eta (W^2,Q^2) \gg \xi$,
the longitudinal part of the photoabsorption cross section in (\ref{2.33a})
becomes more strongly suppressed than the transverse part.

The total photoabsorption cross section, according to (\ref{2.33a}) 
is given by
\bqa
\sigma_{\gamma^*p} (W^2,Q^2) & = & \sigma_{\gamma^*_Lp} (W^2,Q^2) +
\sigma_{\gamma^*_Tp} (W^2,Q^2) = \nonumber \\
& = & \frac{\alpha R_{e^+e^-}}{3 \pi} 
\sigma^{(\infty)} (W^2) 
\left( I^{(1)}_T \left( \frac{\eta}{\rho}, \frac{\mu}{\rho} \right)
G_T (u) + I^{(1)}_L
(\eta, \mu) G_L (u)
\right).\nonumber\\
&&
\label{2.35}
\eqa

%\bqa
%sigma_{\gamma^*p} (W^2,Q^2) & = & \frac{\alpha R_{e^+e^-}}{3 \pi}
%\sigma^{(\infty)} (W^2) \rho I_0 (\eta) \left( \frac{1}{3} G_L (u) + \frac{2}{3}
%G_T (u) \right) \times \nonumber\\
%& \times & \left\{ \begin{array}{l@{\quad,\quad}l}
%(1+0 (u)) & (u \ll 1,~\eta > \xi), \\
%\left( 1 + 0 \left( \frac{1}{u} \right)\right) & (u \gg 1, \eta < 1 \ll \xi),
%\end{array} \right.%\nonumber
%\label{2.36}
%\eqa
%\be
%\frac{1}{3} G_L (u) + \frac{2}{3} G_T (u) = \frac{u}{1+u} =
%\left\{ \begin{array}{l@{\quad,\quad}l}
%u + 0 (u^2) & (u \ll 1), \\
%1 + 0 \left( \frac{1}{u} \right) & (u \gg 1).
%\end{array} \right.
%\label{2.37}
%\ee
%\bqa
%\sigma_{\gamma^*p} (W^2,Q^2) & = & \frac{\alpha R_{e^+e^-}}{3 \pi}
%\sigma^{(\infty)} (W^2) \rho I_0 (\eta) \left( \frac{1}{3} G_L (u) + \frac{2}{3}
%G_T (u) \right) \times \nonumber \\
%& \times & \left( 1 + \frac{1 - \rho}{\rho} \frac{G_L (u)}{G_L (u) + 2 G_T (u)}
%\right),~~~~~(\eta \gg 1).
%\label{2.38}
%\eqa
%\bqa
%G_T(u) & = & \frac{1}{3} \left( G_L (u) + 2 G_T (u) \right) + \frac{1}{3}
%\left( G_T (u) - G_L (u) \right) \nonumber \\
%& = & \frac{1}{3} \left( G_L (u) + 2 G_T (u) \right) + 0 \left( \frac{1}{u}
%\right),~~~~(u \gg 1), 
%\label{2.39}
%\eqa

In the photoproduction limit of (\ref{2.10}), the longitudinal contribution
to (\ref{2.35}) goes to zero, the transverse contribution becomes proportional
to $\ln (\rho/\mu (W^2)) = \ln (\rho \Lambda^2_{sat} (W^2)/m^2_0)$, and
moreover $G_T (u) \simeq 1$ from (\ref{2.34}). Requiring consistency of
(\ref{2.35}) in the limit of $Q^2 \to 0$ with the empirically known 
photoproduction cross section, allows one to determine the hadronic
dipole cross section $\sigma^{(\infty)} (W^2)$. From the $Q^2 \to 0$
limit of (\ref{2.35}) we obtain
\be
\sigma^{(\infty)} (W^2) = \frac{3 \pi}{\alpha R_{e^+e^-}} ~~
\frac{1}{\lim\limits_{\eta \to \mu (W^2)} I_T^{(1)} \left( \frac{\eta}{\rho},~ 
\frac{\mu (W^2)}{\rho}\right)}~~\sigma_{\gamma p} (W^2) .
\label{2.40}
\ee

For the evaluation of $\sigma^{(\infty)} (W^2)$ the Regge fit to the
experimental data for photoproduction,
$\sigma_{\gamma p} (W^2)$, \cite{DIFF2000}, \cite{Donnachie}, or the 
double-logarithmic fit from the
Particle Data Group \cite{PDG},
%\newpage
\bqa
\sigma^{(a)}_{\gamma p} (W^2) & = & 0.0635 (W^2)^{0.097} + 0.145 (W^2)^{-0.5},
\label{2.41} \\
\sigma^{(b)}_{\gamma p} (W^2) & = & 0.0677 (W^2)^{0.0808} 
+ 0.129 (W^2)^{-0.4525}, \nonumber \\
\sigma^{(c)}_{\gamma p} (W^2) & = & 0.003056 \left(33.71 + \frac{\pi}{M^2}
\ln^2 \frac{W^2}{(M_p + M)^2} \right) + 0.0128 \left( \frac{(M_p+M)^2}{W^2}
\right)^{0.462},\nonumber
\eqa
(where $M_p$ stands for the proton mass and $M \equiv 2.15 GeV$) are to
be inserted into (\ref{2.40}).

The $(\ln W^2)^2$ dependence of $\sigma^{(c)}_{\gamma p} (W^2)$
in (\ref{2.41}) together with (\ref{2.27}) implies a growth of $\sigma^{(\infty)} (W^2)$ in 
(\ref{2.40}) 
as $\sigma^{(\infty)} (W^2) \sim \ln W^2$. The photoabsorption
cross section (\ref{2.35}) for fixed $Q^2 \ge 0$, taking into account
(\ref{2.26}) as well as 
$I^{(1)}_T  \sim \ln W^2$ from (\ref{2.27}), then grows as $(\ln W^2)^2$.
The photoabsorption cross section  at any fixed $Q^2 \ge 0$ for $W^2 \to
\infty$   
behaves hadronlike, compare also (\ref{2.28}) \cite{DIFF2000}, \cite{1108}.

The growth of the photoabsorption cross section with increasing energy
as $(\ln W^2)^2$, for any fixed $Q^2 \ge 0$, coincides
with the hadronic $(\ln W^2)^2$ behavior conjectured by Heisenberg
\cite{Heisenberg} for cross sections among strongly interacting 
particles,
and recognized as the maximal possible growth with
energy by Froissart \cite{Froissart}. We note that a hadronlike
Froissart-like saturation behavior of $\sigma_{\gamma^*p} (W^2,Q^2)$
for $W^2$ sufficiently large, $Q^2 \ge 0$ fixed, was recently also 
demonstrated \cite{Block2} by the success of an explicit
``Froissart-inspired'' fit to the DIS experimental data for $x \cong
\frac{Q^2}{W^2} \le 0.1$.

We turn to a discussion of the longitudinal-to-transverse ratio
$R(W^2,Q^2)$. According to (\ref{2.33a}), it is given
by
\be
R(W^2,Q^2)  =  \frac{\sigma_{\gamma^*_Lp} (W^2,Q^2)}{\sigma_{\gamma^*_Tp}
(W^2,Q^2)} 
=  \frac{I_L^{(1)} (\eta,\mu)}{I_T^{(1)} 
\left(\frac{\eta}{\rho},\frac{\mu}{\rho}\right)}
\frac{G_L (u)}{G_T (u)}. \\
\label{2.42}
\ee
In the $Q^2 = 0$ photoproduction limit of $\eta (W^2,Q^2 = 0) 
= \mu (W^2) \ll 1$
(as a consequence of electromagnetic gauge invariance), the ratio
$R(W^2,Q^2)$ becomes zero. For $\eta (W^2,Q^2) \gg 1$, according to
(\ref{2.20}) and (\ref{2.21}),
\be
R(W^2,Q^2) = \frac{1}{2 \rho} \frac{G_L (u)}{G_T (u)},~~~
(\eta (W^2,Q^2) \gg 1).
\label{2.43}
\ee
The (necessary) restriction (\ref{2.30}) on the mass, $M_{q \bar q}$,
of $q \bar q$ fluctuations modifies (\ref{2.24}) to become (\ref{2.43}).
In the region of $u(\eta (W^2,Q^2)) = \xi/\eta (W^2,Q^2) \gg 1$,
according to (\ref{2.34}), we have $G_L (u)/G_T (u) \cong 1$ in
(\ref{2.43}). For $u (\eta(W^2,Q^2)) \ll 1$, we have $G_L (u)/G_T(u) \simeq
2u = 2 \xi/\eta$ corresponding to a strong decrease of $R(W^2,Q^2)$ for 
sufficiently large $Q^2$ at fixed energy $W$.

A measurement of $R (W^2,Q^2)$ for $Q^2$ sufficiently large determines
the magnitude of $\rho$. The parameter $\rho$, according to (\ref{A42})
to (\ref{A44}),
is to be identified with the ratio of the cross sections for transversely
polarized, $(q \bar q)^{J=1}_T$, and longitudinally polarized, 
$(q \bar q)^{J=1}_L$, dipole states
in the limit of vanishing dipole size,
$\vec r^{~\prime 2}_\bot \to 0$,
\be
\rho = \frac{\bar \sigma_{(q \bar q)^{J=1}_T p} (\vec r^{~\prime 2}_\bot, W^2)}
{\bar \sigma_{(q \bar q)^{J=1}_L p} (\vec r^{~\prime 2}_\bot, W^2)}
\Bigg|_{\vec r^{~\prime 2}_\bot \to 0}.
\label{2.44}
\ee
The proportionality of the dipole cross sections in (\ref{2.44}) to the
(transverse) size $\vec r^{~\prime 2}_\bot$ of the $(q \bar q)^{J=1}_{L,T}$
dipole states implies that $\rho$ is independent of the energy $W$ of the
$(q \bar q)p$ interaction.

According to  (\ref{A39}) to (\ref{A44}), 
transverse $q \bar q$ states,
$(q \bar q)^{J=1}_T$, (for $\rho > 1$) interact with enhanced transverse
size compared with $(q \bar q)^{J=1}_L$ states,
\be
\bar\sigma_{(q \bar q)^{J=1}_T  p} (\vec r^{~\prime 2}_\bot, W^2) = 
\bar\sigma_{(q \bar q)^{J=1}_L  p} (\rho \vec r^{~\prime 2}_\bot, W^2), 
~~(\vec r^{~\prime 2}_\bot \to 0).
\label{2.45}
\ee
The parameter $\rho$ directly measures the effect of the enhanced transverse
size of transversely relative to longitudinally polarized $q \bar q$ states.
The prediction \cite{Ku-Schi}, \cite{PRD85} of
\be
\rho = \frac{4}{3}
\label{2.46}
\ee
is based on the assumption that the effect of the enhanced transverse
size can be fully taken care of by employing the average transverse
sizes of $(q \bar q)^{J=1}_T$ and $(q \bar q)^{J=1}_L$ states, when
predicting the relative magnitude of their cross sections. A deviation
from the prediction (\ref{2.46}) is to be interpreted as a dependence
of the interaction of $q \bar q$ states with the proton that is not fully
taken care of by employing the average sizes of the $q \bar q$ states.
A deviation from (\ref{2.46}) thus indicates an additional dependence on
the $z(1-z)$ internal configuration of $(q \bar q)^{J=1}$ states
not incorporated in the averaging 
procedure leading to $\rho = 4/3$.

According to the preceding discussion it is clear that the general 
formulation of the CDP in (\ref{2.1}) and (\ref{2.2}) implies $\rho = const$
due to invariance under Lorentz boosts. Deviations from the prediction 
(\ref{2.46}) for the numerical value of $\rho = 4/3$ from (\ref{2.46}) cannot
be strictly excluded.
In ref. \cite{PRD85},
we introduced a more general ansatz for the dipole cross section
that leads to a parameter-dependent
expression for $\rho = \rho (\epsilon \equiv 1/6a)$. According to (\ref{A46}),
we have \cite{PRD85}
\be
\rho (\epsilon \equiv 1/6a) = \left\{ \begin{array}{r@{\quad{\rm for}\quad}l}
1 & a = 2.54, \\
\frac{4}{3} & a = 5.53, \\
2 & a = 23.2.
\end{array} \right. 
\label{2.47}
\ee
The general ansatz from ref. \cite{PRD85} thus contains helicity independence,
$\rho = 1$, as well as the required enhanced-transverse-size effect of 
$\rho > 1$. 

\section{Theory versus experiment}
\renewcommand{\theequation}{\arabic{section}.\arabic{equation}}
\renewcommand{\thefigure}{\arabic{section}.\arabic{figure}}
\setcounter{equation}{0}

In the present section we present the theoretical results from Section 2
in a form that is convenient to be employed in a fit to the experimental
data. In particular, we present the theoretical results in terms of the
so-called reduced cross section employed in the recent combined analysis 
of the H1- and ZEUS collaborations \cite{H1-ZEUS}. We also elaborate on 
how the photoproduction limit should be tested by the experimental data,
and we present a comparison of the longitudinal-to-transverse ratio
$R(W^2,Q^2)$ with experimental data.

%\be
%\sigma_{\gamma^*p} (W^2,Q^2) = 
%\frac{\sigma_{\gamma p} (W^2)}{\ln \frac{\Lambda^2_{sat} (W^2)}{m^2_0}}
%I_0 (\eta) \frac{u}{1+u},
%\label{3.1}
%\ee

According to (\ref{2.35}) and (\ref{2.40}), 
we have 
%\be
%\sigma_{\gamma^*p} (W^2, Q^2) = \frac{\sigma_{\gamma p} (W^2)}{\ln
%\frac{\Lambda^2_{sat} (W^2)}{m^2_0}} \left( (I_0 (\eta) - I_L^{(1)} 
%(\eta, \mu))G_T (u) + \frac{1}{\rho} I_L^{(1)} (\eta, \mu) G_L (u) \right),
%\label{3.2}
%\ee
\be
\sigma_{\gamma^*p} (W^2, Q^2) = \frac{\sigma_{\gamma p} (W^2)}
{\lim\limits_{\eta \to
\mu (W^2)} I_T^{(1)} 
\left( \frac{\eta}{\rho},
\frac{\mu (W^2)}{\rho}\right)}
 \left( I^{(1)}_T \left(
\frac{\eta}{\rho}, \frac{\mu}{\rho} \right) G_T (u) + I^{(1)}_L (\eta, \mu)
G_L (u) \right).
\label{3.2}
\ee
$I_{L,T}^{(1)} (\eta, \mu)$ and $G_{L,T} (u)$ are given in (\ref{2.9}) or
(\ref{2.14}) and (\ref{2.34}), respectively. According to (\ref{2.9})
\be
\lim_{\eta \to \mu (W^2)} I^{(1)}_T \left( \frac{\eta}{\rho},
\frac{\mu (W^2)}{\rho} \right) = \ln \frac{\rho}{\mu (W^2)},
\label{3.2a}
\ee
while according to (\ref{2.14})
\bqa
\lim_{\eta \to \mu (W^2)} && I^{(1)}_T \left( \frac{\eta}{\rho}, 
\frac{\mu (W^2)}{\rho} \right) = I_0 \left(
\frac{\mu (W^2)}{\rho} \right) \nonumber \\
&& \cong \ln \frac{\rho}{\mu (W^2)} \left( 1 - 2 \frac{\mu}{\rho} 
\left( 1 - \frac{1}{\ln \frac{\rho}{\mu}} \right)\right) < \ln \frac{\rho}
{\mu (W^2)}.
\label{3.2b}
\eqa
For the photoproduction cross section, $\sigma_{\gamma p} (W^2)$, in
(\ref{3.2}) the empirical fit in (\ref{2.41}) is to be inserted. Independently
of whether (\ref{2.9}) or (\ref{2.14}) is employed in (\ref{3.2}), we have
convergence to the photoproduction $\lim\limits_{Q^2 \to 0} \sigma_{\gamma^*p}
(W^2,Q^2) = \sigma_{\gamma p} (W^2)$ in (\ref{3.2}).The somewhat smaller
value of $I^{(1)}_T (\eta, \mu)$ in the limit of $\eta \to \mu$ in (\ref{3.2b})
compared with (\ref{3.2a}) at $\eta \gg 1$ leads to a slightly
larger cross section (\ref{3.2}) for the case
of (\ref{2.14}) that may be absorbed into a somewhat smaller value of $\xi$ in
$m^2_1 (W^2) = \xi \Lambda^2_{sat} (W^2)$.

The longitudinal-to-transverse ratio $R(W^2,Q^2)$, according to (\ref{2.42})
as well as (\ref{3.2}), becomes
%\bqa
%R(W^2,Q^2) & = & \frac{\sigma_{\gamma^*_Lp} (W^2,Q^2)}{\sigma_{\gamma^*_Tp}
%(W^2,Q^2)} 
%=  \frac{1}{\rho} \frac{I_L^{(1)} (\eta,\mu)}{I_T^{(1)} (\eta,\mu)}
%\frac{G_L (u)}{G_T (u)}, \\
%& = & \frac{1}{\rho} \frac{I_L^{(1)} (\eta,\mu)}{I_0 (\eta) - I_L^{(1)} 
%(\eta, \mu)} \frac{G_L (u)}{G_T(u)}.\nonumber
%\label{3.3}
%\eqa
\be
R(W^2,Q^2)  = \frac{\sigma_{\gamma^*_Lp} (W^2,Q^2)}{\sigma_{\gamma^*_Tp}
(W^2,Q^2)} 
= \frac{I_L^{(1)} (\eta,\mu)}{I_T^{(1)} \left(\frac{\eta}{\rho},
\frac{\mu}{\rho}\right)}
\frac{G_L (u)}{G_T (u)}.
\label{3.3}
\ee
From (\ref{2.34}), one finds that the factor $G_L(u)/G_T(u)$ in (\ref{3.3})
is given by
\be
\frac{G_L(u)}{G_T(u)} \cong \left\{ \begin{array}{r@{\quad,\quad}l}
1 + \frac{3}{2u} = 1 + \frac{3}{2} \frac{\eta (W^2,Q^2)}{\xi} & \mbox{for $\eta
  \ll \xi$}, \\
2u = 2 \frac{\xi}{\eta (W^2,Q^2)}\hfill & \mbox{for $\eta \gg \xi$}.
\end{array} \right.
\label{3.3a}
\ee
For $Q^2 = 0$ photoproduction, $\eta (W^2, Q^2) = \mu (W^2)$, according to 
(\ref{2.26}) and (\ref{2.27}), the ratio $I_L^{(1)} (\eta, \mu)/I_T^{(1)}
(\eta, \mu)$ in (\ref{3.3}) goes to zero,
\be
R(W^2,Q^2 = 0) = 0.
\label{3.3b}
\ee
For $\eta (W^2,Q^2) \gg 1$, from (\ref{2.20}) and (\ref{2.21}), we find
\be
R (W^2,Q^2) \cong \frac{1}{2 \rho} \frac{G_L (u)}{G_T (u)},~~~(\eta \gg 1),
\label{3.3c}
\ee
and upon substituting (\ref{3.3a}),
\be
R (W^2, Q^2) \cong \frac{1}{2 \rho} \left\{ \begin{array}{r@{\quad,\quad}l}
(1 + \frac{3}{2} \frac{\eta}{\xi}) & \mbox{for $1 \ll \eta \ll \xi$,}\\
\frac{\xi}{\eta}\hfill & \mbox{for $\eta \gg \xi$.}
\end{array} \right.
\label{3.3d}
\ee
The ratio $R (W^2, Q^2)$, for e.g. fixed energy $W$, with increasing
$\eta \cong  Q^2/\Lambda^2_{sat} (W^2)$ increases proportional to $Q^2$,
and upon reaching a maximum decreases inversely proportional to $Q^2$
as $1/\eta \cong \Lambda^2_{sat} (W^2)/Q^2$. In the approximation of ignoring
the finiteness of $m^2_1 (W^2)$, for $\xi \to \infty$, we have
$R \cong 1/2 \rho$ for sufficiently large $\eta (W^2, Q^2) \gg \mu (W^2)$.

The total photoabsorption cross section in (\ref{3.2}) and 
the ratio $R(W^2,Q^2)$ in (\ref{3.3}) depend on the numerical values of
the saturation scale $\Lambda^2_{sat} (W^2)$,
and the lower and upper bounds, $m^2_0$ and $m^2_1 (W^2) = \xi \Lambda^2_{sat}
(W^2)$, on the masses of the $q \bar q$ fluctuations. The saturation scale 
is determined by its normalization $C_1$ and the exponent $C_2$,
\be
\Lambda^2_{sat} (W^2) = C_1 \left( \frac{W^2}{1 GeV^2} \right)^
{\displaystyle C_2}.
\label{3.4}
\ee
Compare also the previously employed \cite{DIFF2000} form
\be
\Lambda^2_{sat}(W^2) = B \left(1 + \left( \frac{W^2}{W^2_0} \right)
\right)^{\displaystyle C_2} 
\cong C_1 \left( \frac{W^2}{1 GeV^2} \right)^{\displaystyle C_2}.
\label{3.5}
\ee
The successful representation of the experimental data in \cite{PRD85} was based on
\bqa
B & = & 2.04 GeV^2 ,\nonumber \\
W^2_0 & = & 1081 GeV^2, \nonumber \\
C_2 & = & 0.27 ,\nonumber \\
m^2_0 & = & 0.15 GeV^2 , \nonumber\\
\xi & = & 130.
\label{3.6}
\eqa
The parameter $\rho$, from an estimate \cite{Ku-Schi, PRD85}  based on the uncertainty relation
as applied to the different $z(1-z)$ configurations of $q \bar q$ states from
$\gamma^*_L \to q \bar q$ and $\gamma^*_T \to q \bar q$, was determined
to be $\rho = 4/3$.

In fig. 3.1, we compare the results from the simple closed expression for the
photoabsorption cross section in (\ref{3.2}) with the previous more involved
form evaluated in \cite{PRD85}, 
and we find good agreement. The evaluation of the cross section (\ref{3.2})
was carried out with the numerical values of the parameters given in
(\ref{3.6}).
For easy reference, in fig. 3.2, we reproduce 
the comparison with experiment shown in fig. 9 of ref. \cite{PRD85}.

\begin{figure}
%\vspace*{-5cm}
\begin{minipage}[t]{8cm}
\epsfig{file=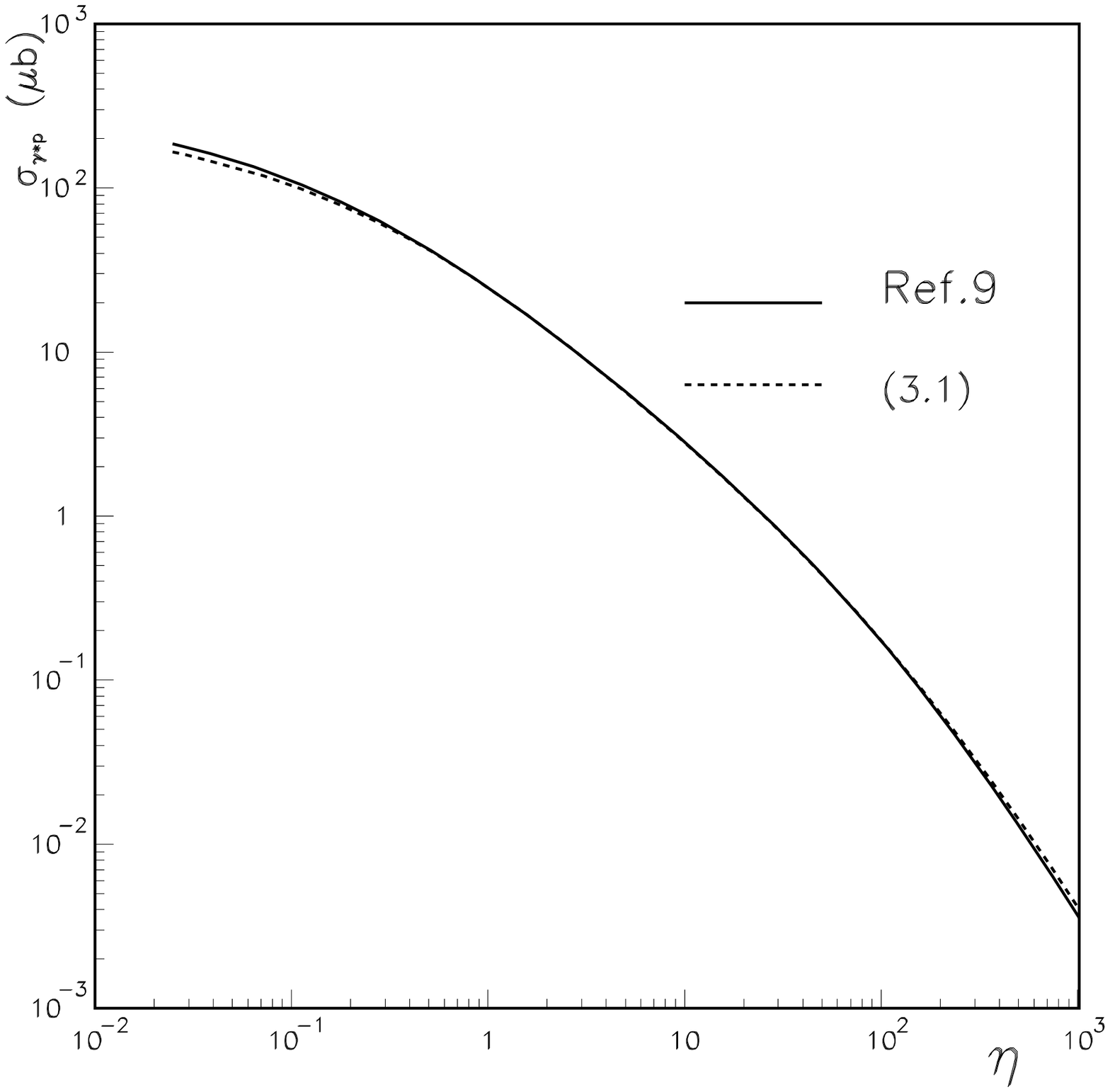,width=7.9cm} 
\caption{\label{fig3.1}
Comparison of the photoabsorption cross section based on
the simple closed analytic expression (\ref{3.2}) with the result from
ref. \cite{PRD85}}
\end{minipage}
\hspace{0.3cm}
%\vspace*{1cm}
\begin{minipage}[t]{8.5cm}
%\vspace*{0.1cm}
\epsfig{file=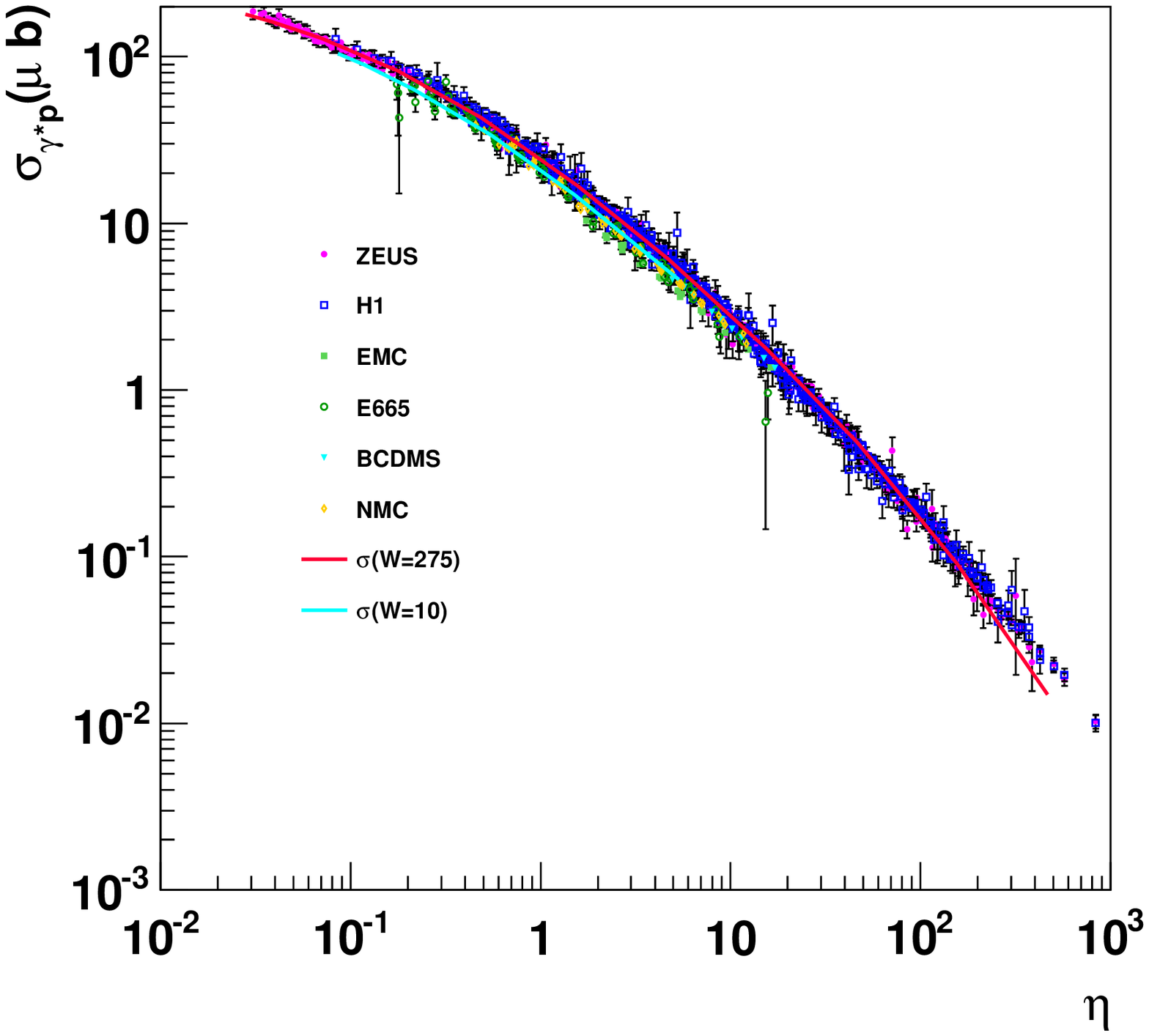,width=8.5cm} 
\caption{\label{fig3.2}
Comparison of the photoabsorption cross section with experiment,
as shown in ref. \cite{PRD85}}
\end{minipage}
\end{figure}

\begin{figure}
%\vspace*{-5cm}
\centerline{\epsfig{file=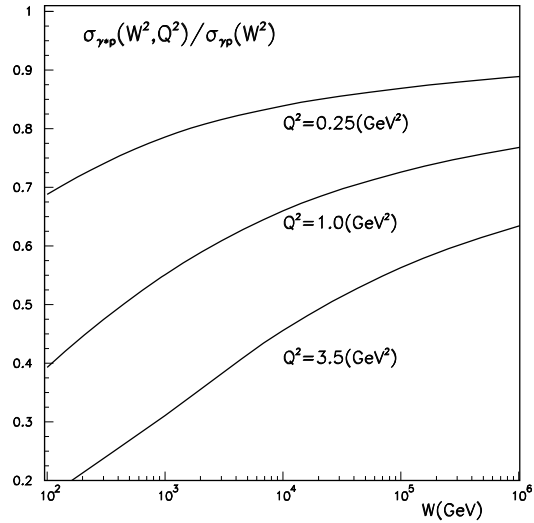,width=7.9cm} }
\caption{\label{fig3.3}
The $W$ dependence of the ratio (\ref{2.28}) of $\sigma_{\mu p}
(W^2,Q^2)/\sigma_{\gamma p} (W^2)$ for several values of $Q^2$ showing
the slow convergence to the hadronlike dependence of $\sigma_{\gamma p}
(W^2) \sim (\ln W^2)^2$ for any fixed value of $Q^2 > 0$.}
\end{figure}

The results shown in fig. 3.3 are relevant for the asymptotic behavior
\cite{DIFF2000}, \cite{1108}, \cite{PRD85} of
the photoabsorption cross section, 
$\sigma_{\gamma^*p} (W^2,Q^2)/\sigma_{\gamma p} (W^2) \to 1$ for
$W^2 \to \infty$, as given by (\ref{2.28}). Dividing (\ref{3.2})
by the $Q^2 = 0$ photoproduction cross section, we show the ratio of
$\sigma_{\gamma^*p} (W^2,Q^2)/\sigma_{\gamma p} (W^2)$ as a function of $W$ 
for various values of $Q^2$. The figure illustrates the very slow
approach to the saturation limit of photoproduction. The slow approach to this
unique limit of the theoretically well-founded CDP
differs strongly from the results of an {\it ad hoc} 
fit\footnote{Neither the underlying 
approximation of 
$\tau \approx 1/x$, ignoring $M^2_{q \bar q}$ in (\ref{2.29}) employed in
\cite{Caldwell}, nor the assumed power-law dependence of $\sigma_{\gamma^* p}
\sim (1/x)^{\lambda (Q^2)}$ is based on theoretical principles.} presented
more recently in ref. \cite{Caldwell}. The high-energy extrapolation of
this ad hoc fit at fixed $Q^2$ leads to a break down of the fit
(``crossing'') for values of $x$ around $x \simeq 10^{-8}$ corresponding
to $W \simeq 10^4~ GeV$ to $W \simeq 10^5~GeV$.

The H1 and ZEUS Collaborations have recently presented \cite{H1-ZEUS}
a combined analysis of their measurements on electron (positron)-proton
scattering carried out at the ep collider HERA from 1992 to 2007.
The results of the experiment are given in terms of the so-called
reduced cross sections \cite{H1-ZEUS}
\be
\sigma^\pm_{r, NC} = \frac{d^2 \sigma^{e^\pm p}_{NC}}{dx_{Bj} dQ^2}
\frac{Q^4 x_{Bj}}{2 \pi \alpha^2 Y_+},
\label{3.12}
\ee
where NC refers to the neutral-current electron (positron) scattering
process. In the case of $Q^2 \ll M^2_Z$ we are concerned with in the
present note, the reduced cross section is related to the proton
electromagnetic structure functions $F_2(x,Q^2)$ and $F_L(x,Q^2)$ via
\bqa
\sigma^\pm_{r, NC} &&\equiv \sigma_r (Q^2, x_{bj}, s) \nonumber \\
&& = F_2 (x,Q^2) - \frac{y^2}{1+(1-y)^2} F_L (x,Q^2),
\label{3.13}
\eqa
where $x \equiv x_{bj},~Q^2 = - q^2 > 0$ with $q^2$ being the
four-momentum transfer from the electron (positron) to the proton, and 
$y$ denotes the ratio of the hadronic center-of-mass energy squared,
$W^2$, to the total $e^\pm p$ energy squared, $s$. In detail,
\be
x_{bj} \equiv x = \frac{Q^2}{W^2+Q^2-M^2_p} \cong \frac{Q^2}{W^2},
\label{3.14}
\ee
and
\be
y = \frac{Q^2}{sx} \cong \frac{W^2}{s},~~~~~~ Y_+ = 1 + (1-y)^2.
\label{3.15}
\ee

The precision experimental data \cite{H1-ZEUS} call for a detailed
examination of the low $x$ region, $x \lsim 0.1$, and in particular for
a close examination of the important $Q^2 \to 0$ photoproduction limit
that is excluded in the analysis of reference \cite{H1-ZEUS}.

The pair of variables relevant at low values of $x \lsim 0.1$ is
the pair $(Q^2,W^2)$, since the transition to $Q^2 = 0$ photoproduction 
requires $Q^2 \to 0$ at fixed $W^2$.\footnote{The limit $Q^2 \to 0$
at fixed $W^2$ in the pair of variables $(Q^2,x)$ corresponds to the
(less convenient) transition of $Q^2 \to 0$ at fixed values of
$Q^2/x$, where $Q^2/x = W^2 - M^2_p = {\rm const}$ is to be required.}
It is also appropriate to replace the structure functions $F_2 (x,Q^2)$
and $F_L (x,Q^2)$ by the photoabsorption cross sections for 
transversely and longitudinally polarized (virtual) photons \cite{Devenish}
$
\sigma_{\gamma^*_Tp} (W^2,Q^2)~ \mbox{and}~ \sigma_{\gamma^*_Lp}
(W^2,Q^2),
$
via
\be
F_2 = \frac{Q^2}{4 \pi^2 \alpha} \left( \sigma_{\gamma^*_Tp}
(W^2,Q^2) + \sigma_{\gamma^*_L p} (W^2,Q^2) \right),
\label{3.16}
\ee
and
\be
F_L = \frac{Q^2}{4 \pi^2 \alpha} \sigma_{\gamma^*_Lp} (W^2,Q^2),
\label{3.17}
\ee
where, for the presently relevant case of $W^2 \gg Q^2$ as well as
$W^2 \gg M^2_p$ (with $M_p$ denoting the proton mass), in (\ref{3.16})
and (\ref{3.17}) the approximation
\be
\frac{Q^4 (1-x)}{4 \pi^2 \alpha (Q^2 + (2M_px)^2)} \cong
\frac{Q^2}{4 \pi^2 \alpha}
\label{3.18}
\ee
was inserted. Replacing the structure functions in (\ref{3.13}) by the
photoabsorption cross sections according to (\ref{3.16}) and (\ref{3.17}), with
\be
\sigma_{\gamma^*p} (W^2,Q^2) \equiv \sigma_{\gamma^*_Tp} +
\sigma_{\gamma^*_Lp} (W^2,Q^2)
\label{3.19}
\ee
and
\be
R(W^2,Q^2) \equiv \frac{\sigma_{\gamma^*_Lp} (W^2,Q^2)}{\sigma_{\gamma^*_Tp}
(W^2,Q^2)},
\label{3.20}
\ee
we obtain
\bqa
\frac{4 \pi^2 \alpha}{Q^2} \sigma_r (W^2, Q^2,s) & = &
\sigma_{\gamma^*_Tp} (W^2,Q^2) \left(1+\left(1 - \frac{y^2}{1+(1-y)^2} \right)
R (W^2,Q^2) \right) \nonumber \\
& = & \sigma_{\gamma^*p} (W^2,Q^2) \left(1 + \frac{y^2}{1 + (1-y)^2}
\frac{R(W^2,Q^2)}{1 + R(W^2,Q^2)} \right).
\label{3.21}
\eqa
Since $y^2$ and $R(W^2,Q^2)$ are small compared
with unity, the main contributions from the right-hand side in (\ref{3.21})
are due to the first terms in the brackets. With $R(W^2,Q^2) \to 0$ as 
a consequence of electromagnetic gauge invariance,  
photoproduction, $\sigma_{\gamma p}(W^2)$ from (\ref{3.21}), is obtained via
\be
\lim_{Q^2 \to 0 \atop W^2 {\rm fixed}} \frac{4 \pi^2 \alpha}{Q^2}
\sigma_r (W^2,Q^2,s) = \sigma_{\gamma p} (W^2).
\label{3.22}
\ee

The experimental results from reference \cite{H1-ZEUS}, given
for $ep$ energies of $\sqrt s = 318~ \mbox{GeV},$\break 
$300~ \mbox{GeV and}~ 251~
\mbox{GeV}$ in bins of $(Q^2,x)$, have to be converted into bins 
of $(Q^2,W^2)$.The corresponding very elaborate analysis is better carried
out by the experimentalists responsible for these data \cite{H1-ZEUS}.

\begin{figure}
%\vspace*{-5cm}
\begin{minipage}[t]{8cm}
\epsfig{file=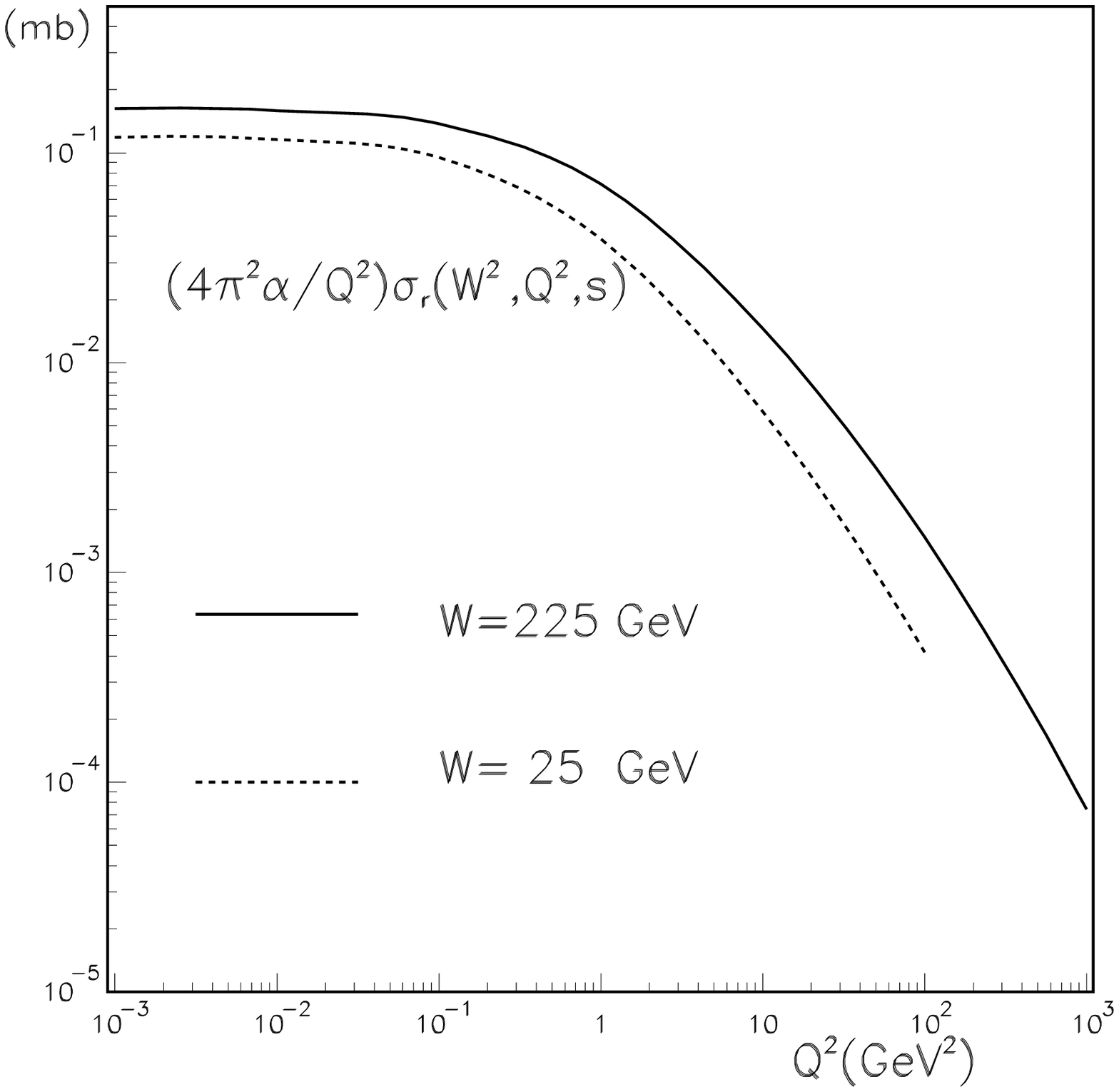,width=7.9cm} 
\caption{\label{fig3.4}
The theoretical results for the reduced cross section
multiplied by $4 \pi^2 \alpha/Q^2$, compare (\ref{3.19}) to (\ref{3.21}), as
a function of $Q^2$ for fixed $W$ at $\sqrt{s} = 318~\mbox{GeV}$.}
\end{minipage}
\hspace{0.3cm}
\begin{minipage}[t]{8cm}
\epsfig{file=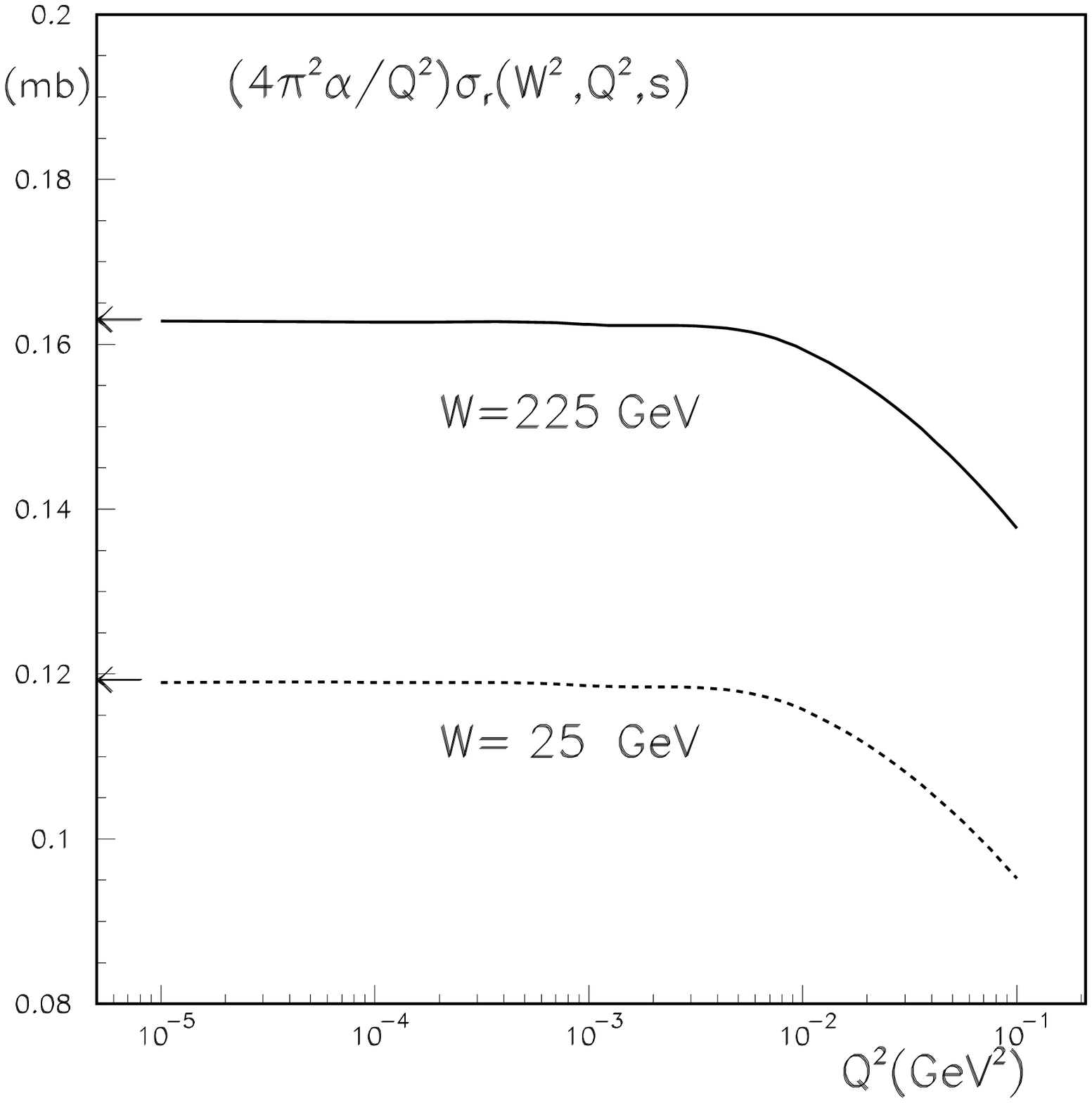,width=7.9cm} 
\caption{\label{fig3.5}
Same as fig. 3.4, but for $Q^2 < 0.1~\mbox{GeV}^2$}
\end{minipage}
\end{figure}

In figs. 3.4 and 3.5, we present the theoretical results for the reduced
cross section at $\sqrt{s} = 318 \mbox{GeV}$
multiplied by $4 \pi^2 \alpha/Q^2$, compare (\ref{3.21}). The photoabsorption
cross section and the ratio $R(W^2,Q^2)$ entering (\ref{3.21}) are
obtained by evaluating (\ref{3.2}) and (\ref{3.3}) with the parameters
given in (\ref{3.6}) and used for figs. 3.1 to 3.3. More specifially, in
fig. 3.4 we show the theoretical results for the cross 
section (\ref{3.21}) for fixed
$W^2$ as a function of $Q^2$,  $Q^2$ being restricted by $x \cong
Q^2/W^2 < 0.1$. In fig. 3.5, we concentrate on the region of very small $Q^2$,
in order to examine the $Q^2 = 0$ limit of photoproduction in more detail.

\begin{figure}
%\vspace*{-5cm}
\begin{minipage}[t]{8cm}
\epsfig{file=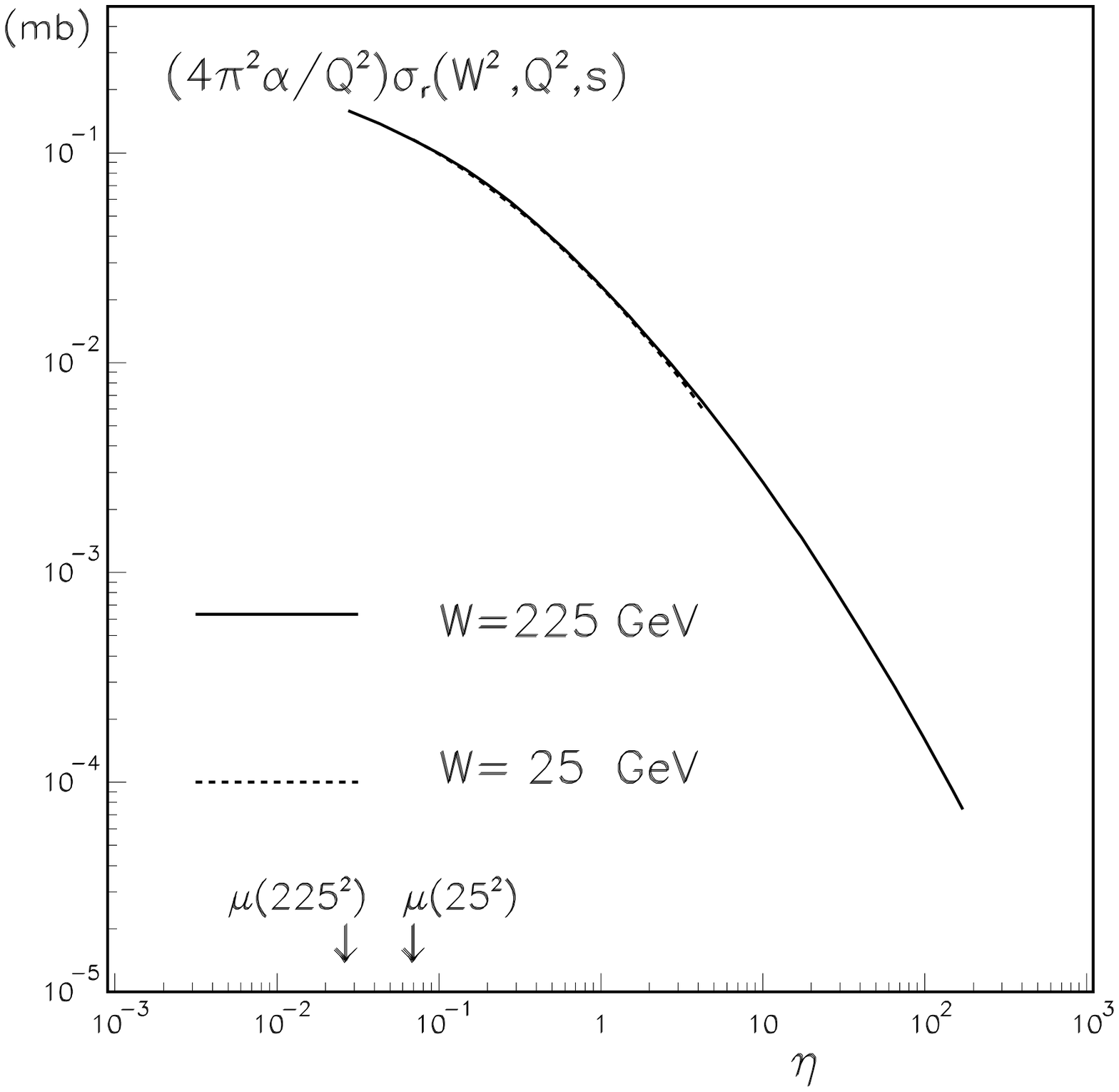,width=7.9cm} 
\caption{\label{fig3.6}
Scaling of $(4 \pi^2 \alpha/Q^2) \sigma^r (W^2,Q^2)$ in the
low-x scaling variable $\eta (W^2,Q^2)$, compare (\ref{3.23}) as
well as (\ref{3.26}) and (\ref{3.21}). As in figs. 3.4 and 3.5, 
$\sqrt s = 318 GeV$.}
\end{minipage}
\hspace{0.3cm}
\begin{minipage}[t]{8cm}
\epsfig{file=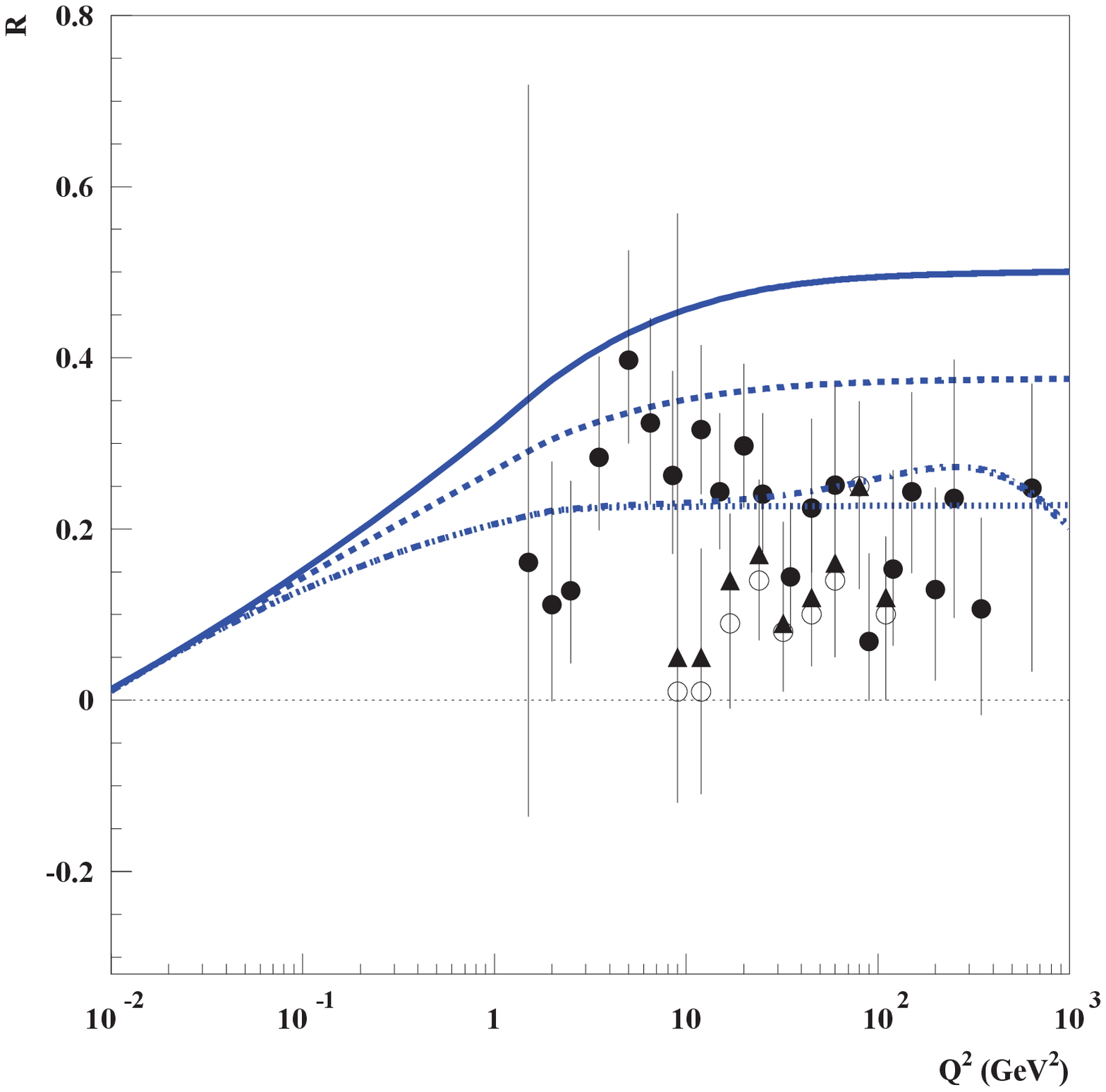,width=7.5cm} 
\caption{\label{fig3.7}
The experimental values of $R (W^2, Q^2)$ at $W \simeq 200~\mbox{GeV}$
compared with the prediction from the CDP.The curves correspond to $\rho = 1$,
$\rho = 4/3$ and $\rho = 2$, respectively.}
\end{minipage}
\end{figure}

In fig. 3.6, we show the theoretical results for the reduced cross section
multiplied by $4 \pi^2 \alpha /Q^2$, compare (\ref{3.21}), as a function
of the low-x scaling variable, (\ref{2.5}) and (\ref{3.4}),
\be
\eta (W^2,Q^2) \equiv \frac{Q^2 + m^2_0}{\Lambda^2_{sat} (W^2)},
\label{3.23}
\ee
where
\be
\Lambda^2_{sat} (W^2) = C_1 
\left(\frac{W^2}{1~ {\rm GeV}^2} \right)^{C_2}
\label{3.24}
\ee
with
%\newpage
\bqa
m^2_0 & = & 0.15~ \mbox{GeV} \nonumber \\
C_1 & = & 0.31 \nonumber \\
C_2 & = & 0.27
\label{3.25}
\eqa
Since the second term in the bracket on the right-hand side in (\ref{3.21})
is small, the prediction \cite{DIFF2000} from the color-dipole
picture (CDP) of low-x scaling,
\be
\sigma_{\gamma^*p} (W^2,Q^2) = \sigma_{\gamma^*p} \left(\eta (W^2,Q^2)\right),
\label{3.26}
\ee
is approximately valid for $\sigma_r~ (Q^2,W^2,s)$. 

We turn to the comparison of our prediction for $R(W^2,Q^2)$,
given in (\ref{3.3b}) to (\ref{3.3d}), with the
H1 and ZEUS experimental results \cite{H1_hep-ex}. 
In a first step, we ignore the upper limit on $m^2_1 (W^2)$
by adopting $\xi \to \infty$. According to (\ref{3.3b}) and
(\ref{3.3d}),
\be
R (W^2, Q^2) \to \left\{ \begin{array}{r@{\quad, \quad}l}
0 & \mbox{for $Q^2 \to 0$}, \\
\frac{1}{2 \rho} & \mbox{for $\eta \simeq \frac{Q^2}{\Lambda^2_{sat} (W^2)}
\gg 1$}.
\end{array} \right.
\label{3.27}
\ee
For the experimental data from H1 and ZEUS belonging to fixed $W$ of
$W \cong {\rm 200~GeV}$, the condition of $\eta (W^2, Q^2) \gg 1$ in
(\ref{3.27}) is fulfilled for $Q^2 \gg 10$. We accordingly predict
\be
R (W^2 \simeq (200~ GeV)^2, Q^2) \to \left\{ \begin{array}{r@{\quad, \quad}l}
0 &\mbox{for $Q^2 \to 0$}, \\
\frac{1}{2 \rho} & \mbox{for $Q^2 \gg 10~ GeV^2$}
\end{array} \right.
\label{3.28}
\ee
i.e. an approach to a constant value of $R \cong 1/2 \rho$ for sufficiently
large $Q^2$. We recall that the factor $1/2$ in (\ref{3.28}) originates
from the ratio of the total $\gamma^* (q \bar q)$ transition strengths for
longitudinally and transversely polarized photons, while $\rho$ stands for
the enhanced cross section for transversely relative to longitudinally
polarized $q \bar q$ states. From an estimate based on the uncertainty
principle, we predicted $\rho = 4/3$ \cite{Ku-Schi, PRD85}.

%\begin{figure}[h]
%\begin{center}
%\vspace*{-5cm}
%\begin{minipage}[t]{8cm}
%\epsfig{file=fig_r.eps ,width=9cm} 
%\caption{\label{fig3.7}
%The experimental values of $R(W^2,Q^2)$ at $W \simeq 200 GeV$
%compared with the prediction from the CDP. The curves correspond to $\rho = 1$,
%$\rho = 4/3$ and $\rho = 2$, respectively.}
%\end{minipage}
%\hspace{0.3cm}
%\begin{minipage}[t]{8cm}
%\epsfig{file=fig2_191115.eps,width=7.9cm} 
%\caption{Same as fig. 1, but for $Q^2 < 0.1~\mbox{GeV}^2$}
%\end{minipage}
%\end{center}
%\end{figure}

The experimental results in fig.\ref{fig3.7}\footnote{Figure 3.7 was
prepared by B. Surrow.}  are 
consistent\footnote{In Fig. \ref{fig3.7}, we 
also show that the result of taking into account $\xi = 130$ is negligible
in the presently investigated kinematic saturation.} with the prediction
(\ref{3.28}). Previous experimental data \cite{PRD85} showed consistency for
$R(W^2, Q^2)$ based on $\rho = 4/3$. The present more accurate experimental
data demonstrate the expected approach to a constant value of $\rho$, but
they require the larger value of $\rho \cong 2$. The transverse-size
enlargement is somewhat larger than the prediction of $\rho = 4/3$ from the
uncertainty principle. 
This indicates that the interaction mechanism of the
$q \bar q$ dipole with the gluon field in the nucleon cannot be fully
reduced to a different value of the average  of the $z(1-z)$ 
distribution (compare (\ref{A39}) and (\ref{A40})),
when passing from longitudinally to transversely polarized $q \bar q$ states.

\section{Conclusion.}
The total photoabsorption cross section is determined by the imaginary part
of the (virtual) Compton-forward-scattering amplitude. At low values of
the Bjorken scaling variable $x \lsim 0.1$, the imaginary part of the
forward-scattering amplitude factorizes into a transition of the
photon to quark-antiquark pairs and their subsequent forward scattering
in the gluon field of the nucleon. The gauge-invariant coupling of the
quark-antiquark color dipole to gluons, at any fixed center-of-mass
energy $W$, implies a hadron-like energy dependence of the cross section
at small values of the photon virtuality, and a stronger increase with $W$ at
large values of the photon virtuality, where the dipole acts as a color-neutral
state. The dependence on the relative magnitude of $Q^2$ and $W^2$
implies the observed low-x scaling behavior in terms of the scaling
variable $\eta (W^2,Q^2)$. At any fixed $Q^2$, at sufficiently high
energy $W$, the transition occurs from color transparency to hadron-like
behavior.

It has been the aim of the present paper to reduce the formalism of
the CDP to most simple closed analytic expressions for the 
photoabsorption cross section for transversely and longitudinally
polarized photons. The resulting total cross section agrees with
experimental data. A detailed global analysis of all experimental
data on the basis of our theoretical results is better carried out
by the experimental groups responsible for most of the high-energy
experimental results.

With respect to the longitudinal-to-transverse ratio, $R (W^2,Q^2)$,
our predictions are consistent with the approximate constancy observed
at sufficiently large $Q^2$.
The absolute magnitude of $R (W^2,Q^2) = 1/(2 \rho)$, where the 
factor $\rho$ is due to the enhanced transverse size of transversely
relative to longitudinally polarized dipole states, shows that the
experimental value of $\rho \cong 2$ is somewhat larger than the
estimate from the uncertainty principle of $\rho \cong 4/3$.
This implies that the difference in the cross sections of
transversely-versus-longitudinally polarized dipole states cannot be
fully reduced to the simple picture of average transverse sizes 
without a more detailed consideration of the effect of the different
quark and antiquark transverse momenta in the quark-antiquark dipole
states.

\bigskip

\begin{appendix}

\renewcommand{\theequation}{\Alph{section}.\arabic{equation}}
\setcounter{section}{1}
\setcounter{equation}{0}
\renewcommand{\thefigure}{\Alph{section}.\arabic{figure}}
\setcounter{section}{1}
\setcounter{figure}{0}
\section*{Appendix A. Details on the Derivation of the Photoabsorption Cross
Section (\ref{2.33a}).}
In this Appendix, we provide a brief exposition of the 
derivation\footnote{The present derivation expands and improves
the treatment in Appendix C of ref. \cite{PRD85}.} of the
photoabsorption cross sections (\ref{2.33a}) and (\ref{2.35}) which
incorporate the upper limit $M^2_{q \bar q} \le m^2_1 (W^2)$,
compare (\ref{2.30}), on the mass, $M_{q \bar q}$, of $q \bar q$
fluctuations of the (virtual) photon.

We also elaborate on the introduction of the transverse-size-enhancement
factor $\rho$ in (\ref{2.24}) and (\ref{2.33}).

Upon transition to momentum space, and upon introducing the mass 
variables
\be
M^2 = \frac{\vec k^{~2}_\bot}{z(1-z)},~~~~~ M^{\prime 2} = \frac{(\vec k_\bot
+ \vec l_\bot)^2}{z(1-z)},
\label{A1}
\ee
for $q \bar q$ states, as well as
\be
\vec l^{~\prime 2}_\bot = \frac{\vec l^{~2}_\bot}{z(1-z)},
\label{A2}
\ee
where $\vec k_\bot$ and $\vec l_\bot$ refer to transverse three momenta of
(massless) quarks and gluons, the photoabsorption cross sections in
(\ref{2.1}) with (\ref{2.2}) become \cite{Cvetic, DIFF2000, PRD85}
\bqa
\hspace*{-1cm}
\sigma_{\gamma^*_Lp} (W^2,Q^2) & = & \frac{\alpha R_{e^+e^-}}{3 \pi}
\int d \vec l^{~\prime 2}_\bot \bar \sigma_{(q \bar q)^{J=1}_L p}
(\vec l^{~\prime 2}_\bot, W^2) \nonumber  \\
& \times & \int dM^2 \int dM^{\prime 2} w (M^2, M^{\prime 2}, 
\vec l^{~\prime 2}_\bot) \nonumber\\
& \times & Q^2 \left( \frac{1}{(Q^2 + M^2)^2} -
\frac{1}{(Q^2 + M^2)(Q^2 + M^{\prime 2})} \right) 
\label{A3}
\eqa
and
\bqa
\hspace*{-1cm}
\sigma_{\gamma^*_Tp} (W^2,Q^2) & = & \frac{\alpha R_{e^+e^-}}{3 \pi}
\int d \vec l^{~\prime 2}_\bot \bar \sigma_{(q \bar q)^{J=1}_T p}
(\vec l^{~\prime 2}_\bot, W^2)  \nonumber \\
& \times & \int dM^2 \int dM^{\prime 2} w (M^2, M^{\prime 2}, 
\vec l^{~\prime 2}_\bot) \nonumber \\
& \times & \left( \frac{M^2}{(Q^2 + M^2)^2} 
- \frac{M^2 +M^{\prime 2} -
\vec l^{~\prime 2}_\bot}{2(Q^2 + M^2) (Q^2 + M^{\prime 2})}
 \right).
\label{A4}
\eqa
In the transition from (\ref{2.1}) and (\ref{2.2}) to (\ref{A3}) and
(\ref{A4}), we introduced the cross sections $\bar \sigma_{(q \bar q)^{J=1}_L
  p} (\vec l^{~\prime 2}_\bot, W^2)$ and $\bar \sigma_{(q \bar q)^{J=1}_T
  p} (\vec l^{~\prime 2}_\bot, W^2)$ for the scattering of longitudinally
and transversely polarized $J=1~q \bar q$ (vector) states on the proton.
The quantity $R_{e^+e^-}$ is given by $R_{e^+e^-} = 3 \Sigma_q Q^2_q$, the
sum over $q$ running over the squares of the charges of the actively
contributing quark flavors ($\sum_q Q^2_q = 10/9$ for four active flavors).
The Jacobian $w(M^2, M^{\prime 2}, \vec l^{~\prime 2}_\bot)$ in (\ref{A3})
and (\ref{A4}) is given by \cite{Cvetic, DIFF2000,PRD85}
\be
w(M^2, M^{\prime 2}, \vec l^{~\prime 2}_\bot) = \frac{1}{2 M M^\prime
\sqrt{1 - \cos^2 \phi}} = \frac{1}{2 M \sqrt{\vec l^{~\prime 2}_\bot}
\sqrt{1 - \cos^2 \vartheta}},
\label{A5}
\ee
where $\phi$ denotes the angle between $\vec k_\bot$ and $\vec k_\bot +
\vec l_\bot$, and $\vartheta$ denotes the angle between $\vec k_\bot$
and $\vec l_\bot$.
Since
\be
\cos^2 \phi = \frac{1}{4 M^2M^{\prime 2}} (M^2 + M^{\prime 2} - 
\vec l^{~\prime 2}_\bot)^2
\label{A6}
\ee
in (\ref{A5}) is symmetric under exchange of $M^2$ and $M^{\prime 2}$,
also $w(M^2,M^{\prime 2},\vec l^{~\prime 2}_\bot)$ in (\ref{A5}) is
symmetric under this exchange. 

%Upon carrying out the substitutions
%\be
%\frac{1}{2} \left( \frac{1}{Q^2+M^2} - \frac{1}{Q^2+M^{\prime 2}} \right)^2
%\Longrightarrow \frac{1}{(Q^2+M^2)^2} - \frac{1}{(Q^2+M^2)(Q^2+M^{\prime 2})},
%\label{A7}
%\ee
%and
%\bqa
%&& \frac{1}{2} \left( \frac{M^2}{(Q^2+M^2)^2} + \frac{M^{\prime 2}}
%{(Q^2+M^{\prime 2})^2} - \frac{M^2 + M^{\prime 2} - \vec l^{~\prime 2}_\bot}
%{(Q^2+M^2) (Q^2+M^{\prime 2})} \right) \nonumber \\
%&& \Longrightarrow \frac{M^2}{(Q^2+M^2)^2} - \frac{1}{2}
%\frac{M^2 + M^{\prime 2} - \vec l^{~\prime 2}_\bot}{(Q^2 + M^2) (Q^2+M^{\prime
%    2})},
%\label{A8}
%\eqa
%the cross sections (\ref{A3}) and (\ref{A4}) turn into the (asymmetric) form
%employed in our previous analysis \cite{Cvetic, DIFF2000, PRD85}.

Noting that
\be
M^{\prime 2} (M^2, \vec l^{~\prime 2}_\bot, \cos \vartheta) =
M^2 + \vec l^{~\prime 2}_\bot + 2 M \sqrt{\vec l^{~\prime 2}_\bot}
\cos \vartheta
\label{A9}
\ee
and
\be
\frac{\partial M^{\prime 2} (M^2, \vec l^{~\prime 2}_\bot, \cos \vartheta)}
{\partial \vartheta} =  \frac{-1}{w (M^2,M^{\prime 2}, \vec l^{~\prime
    2}_\bot)}
\label{A10}
\ee
the integrations in (\ref{A3}) and (\ref{A4}) over $dM^2$ and $dM^{\prime 2}$
may be replaced by integrations over $dM^2$ and $d \vartheta$,
\bqa
\int dM^2 \int dM^{\prime 2} w (M^2, M^{\prime 2}, \vec l^{~\prime 2}_\bot) 
& = &
\int^{m^2_1 (W^2)}_{m^2_0} dM^2 \int^\pi_0 d \vartheta -
\int^{(\sqrt{\vec l^{~\prime 2}_\bot} + m_0)^2}_{(\sqrt{\vec l^{~\prime
      2}_\bot} - m_0)^2} dM^2~ \int^\pi_{\vartheta_0 (M^2, 
\vec l^{~\prime 2}_\bot)}
d \vartheta \nonumber \\
& - & \int^{m^2_1 (W^2)}_{(m_1 (W^2) - \sqrt{\vec l^{\prime 2}_\bot})^2}
dM^2 \int^{\vartheta_1 (M^2, \vec l^{~\prime 2}_\bot)}_0 d \vartheta.
\label{A11}
\eqa
The bounds
\be
m^2_0 \le M^2, M^{\prime 2} (M^2, \vec l^{~\prime 2}_\bot, \cos \vartheta) \le
m^2_1 (W^2)
\label{A12}
\ee
lead to three terms, as indicated in (\ref{A11}). The first term in (\ref{A11})
takes care of the bound (\ref{A12})
on $M^2$, and the second and third
terms correct for the restrictions on $\vartheta$ that are ignored in the first
term. According to (\ref{A9}), the angles $\vartheta_0 (M^2, \vec l^{~\prime
  2}_\bot)$ and $\vartheta_1 (M^2, \vec l^{~\prime 2}_\bot)$ in (\ref{A11}) are
obtained from
\be
\cos \vartheta_{0,1} (M^2, \vec l^{~\prime 2}_\bot) = \frac{m^2_{0,1} - M^2 -
\vec l^{~\prime 2}_\bot}{2M \sqrt{\vec l^{~\prime 2}_\bot}},
\label{A13}
\ee
where $m^2_1$ stands for $m^2_1 \equiv m^2_1 (W^2)$.

For a sufficiently restricted range of the integrations over 
$d \vec l^{~\prime 2}_\bot$ in (\ref{A3}) and (\ref{A4}), the $M^{\prime 2}$
correction term, the third term on the right-hand side in (\ref{A11}), vanishes 
in the limit of $m^2_1 (W^2) \to \infty$. More specifically, for the ansatz
(\ref{2.3}), or equivalently,
\be
\bar \sigma_{(q \bar q)^{J=1}_L p} (\vec l^{~\prime 2}_\bot, W^2) = 
\bar \sigma_{(q \bar q)^{J=1}_T p} (\vec l^{~\prime 2}_\bot, W^2) =
\frac{\sigma^{(\infty)} (W^2)}{\pi} \delta (\vec l^{~\prime 2}_\bot -
\Lambda^2_{sat} (W^2),
\label{A14}
\ee
upon ignoring the second term on the right-hand side in (\ref{A11})
for $\mu (W^2) = m^2_0/\Lambda^2_{sat} (W^2)\break \ll 1$,
and for $m^2_1 (W^2) \to
\infty$,
one obtains \cite{Cvetic, DIFF2000}
the photoabsorption cross sections described in (\ref{2.7}) to
(\ref{2.17}) of the main text. Neglecting the second term on the right-hand
side in (\ref{A11}), the $m^2_0$-correction term, will be justified in Appendix
C.

In what follows, we consider the effect of the finite bound $m^2_1 (W^2)$,
according to (\ref{A11}) and (\ref{A12}), on the cross sections (\ref{A3}) 
and (\ref{A4}) upon substitution of (\ref{A14}).

The contribution to the photoabsorption cross sections (\ref{A3}) and
(\ref{A4}) corresponding to the first term on the right-hand side in
(\ref{A11}), the dominant contributions, can be fully evaluated analytically,
\be
\sigma_{\gamma^*_L p}^{dom} (W^2,Q^2) = A(W^2) I_L (\Lambda^2_{sat} (W^2),
Q^2, M^2) \bigg|^{m^2_1 (W^2)}_{m^2_0},
\label{A15}
\ee
and
\be
\sigma_{\gamma^*_T p}^{dom} (W^2,Q^2) = A(W^2) I_T (\Lambda^2_{sat} (W^2),
Q^2, M^2) \bigg|^{m^2_1 (W^2)}_{m^2_0},
\label{A16}
\ee
where by definition
\be
A (W^2) = \frac{\alpha R_{e^+e^-}}{3 \pi} \sigma^{(\infty)} (W^2),
\label{A17}
\ee
and
\be
I_L (\Lambda^2_{sat} (W^2), Q^2, M^2) = 
\frac{-Q^2}{Q^2 + M^2} + Q^2 L_1 (\Lambda^2, Q^2, M^2),
\label{A18}
\ee
as well as
\bqa
I_T (\Lambda^2_{sat} (W^2), Q^2, M^2) & = & - \frac{\Lambda^2}{2} L_1 
(\Lambda^2, Q^2,
M^2) - I_L (\Lambda^2, Q^2, M^2) \nonumber \\
& + & \frac{1}{2} \ln
\frac{Q^2 + M^2}{\sqrt{X (\Lambda^2, Q^2, M^2)} + Q^2 + M^2 - \Lambda^2}.
\label{A19}
\eqa
In (\ref{A18}) and (\ref{A19}),
\bqa
X (\Lambda^2, Q^2, M^2) & = & (M^2 - \Lambda^2 + Q^2)^2 + 4 Q^2 \Lambda^2,
\nonumber \\
L_1 (\Lambda^2, Q^2, M^2) & = & \frac{1}{\sqrt{\Lambda^2(\Lambda^2 + 4 Q^2)}}
\nonumber \\
& \times &\ln \frac{\sqrt{\Lambda^2 (\Lambda^2 + 4Q^2)} 
\sqrt{X(\Lambda^2,Q^2,M^2)} +
\Lambda^2 (3Q^2 - M^2 + \Lambda^2)}{Q^2+M^2}\hspace*{1cm}
\label{A20}
\eqa
and $\Lambda^2$ stands for $\Lambda^2 \equiv \Lambda^2_{sat} (W^2)$ on the
right-hand sides in (\ref{A18}) to (\ref{A20}).

We turn to the correction terms, the second and the third term on the
right-hand side in (A.9).  Upon integrating over $d\vartheta$, one obtains
\bqa
  &&\Delta  \sigma_{\gamma_L^*p}^{(m_0^2)}
(W^2,Q^2)  =-A(W^2)  
        \int_{(\Lambda-m_0)^2}^{(\Lambda+m_0)^2} dM^2
\nonumber \\
  && ~~~\times\Bigl[ {{Q^2}\over{(Q^2+M^2)^2}}{{\pi-\theta_0(\Lambda^2,M^2)}\over \pi}
           \nonumber \\
  &&~~~~~~~-{{Q^2}\over{(Q^2+M^2)\sqrt{X(\Lambda^2,Q^2,M^2)}  }}
          \bigl(1-{2\over \pi}
          {\rm arctan}\sqrt{Y_0(\Lambda^2,Q^2,M^2)}\bigr)\Bigr] ,
\label{K1}
\eqa
and
\bqa
 && \Delta\sigma_{\gamma^*_Tp}^{(m_0^2)}(W^2,Q^2) 
  =-A(W^2){1\over 2}
      \int_{(\Lambda-m_0)^2}^{(\Lambda+m_0)^2} dM^2   \nonumber \\
  && ~~~\times\Bigl[ {{M^2-Q^2}\over{(Q^2+M^2)^2}}{{\pi-\theta_0(\Lambda^2,M^2)}\over \pi}
             \nonumber \\ 
  &&~~~~~~~- {{M^2-Q^2-\Lambda^2}\over{(Q^2+M^2)\sqrt{X(\Lambda^2,Q^2,M^2)} }}
          \bigl(1-{2\over \pi}
          {\rm arctan}\sqrt{Y_0(\Lambda^2,Q^2,M^2)}\bigr)\Bigr], \label{K2}  
\eqa
as well as
\bqa
 && \hspace*{-1.9cm}\Delta \sigma^{(m_1^2)}_{\gamma_L^*p}(W^2,Q^2)
    =- A(W^2)
       \int_{(m_1-\Lambda)^2}^{m_1^2}dM^2   \nonumber \\
 && \hspace*{-1.9cm}\times \Bigl[{{Q^2}\over{(Q^2+M^2)^2}}{{\theta_1(\Lambda^2,M^2)}\over\pi}
            -{{Q^2}\over{(Q^2+M^2)\sqrt{X(\Lambda^2,Q^2,M^2)}}}{2\over\pi}
      {\rm arctan}\sqrt{Y_1(\Lambda^2,Q^2,M^2)}\Bigr],
\label{K3} 
\eqa
and
\bqa
 && \hspace*{-1.9cm}  \Delta \sigma^{(m_1^2)}_{\gamma_T^*p}(W^2,Q^2)
    = - A(W^2) \frac{1}{2}
       \int_{(m_1-\Lambda)^2}^{m_1^2}dM^2 \nonumber \\
    && \hspace*{-1.9cm} \times \Bigl[{{M^2-Q^2}\over{(Q^2+M^2)^2}}{{\theta_1(\Lambda^2,M^2)}\over\pi}
            -{{M^2-Q^2-\Lambda^2}\over{(Q^2+M^2)
         \sqrt{X(\Lambda^2,Q^2,M^2)}}}{2\over\pi}
      {\rm arctan}\sqrt{Y_1(\Lambda^2,Q^2,M^2)} \Bigr],
\label{K4}
\eqa
where $\theta_j(\Lambda^2,M^2)$  and $Y_j(\Lambda^2,Q^2,M^2)$ are given by
\bq
     \theta_j(\Lambda^2,M^2) = {\rm arccos}\Bigl({{m_j^2-M^2-\Lambda^2}\over
               {2\sqrt{M^2\Lambda^2}}}\Bigr),
         ~~~~~(0\le \theta_j \le \pi),
\label{K5}
\eq
and

\bq
    Y_j(\Lambda^2,Q^2,M^2)= {{Q^2+(M-\Lambda)^2}\over{Q^2+(M+\Lambda)^2}}
\times
        {{1-\cos\theta_j(\Lambda^2,M^2)}\over{1+\cos\theta_j(\Lambda^2,M^2)}},
\label{K6}
\eq
where $j=0,1$.

The dependence of $\theta_j(\Lambda^2,M^2)$ and  $Y_j(\Lambda^2,Q^2,M^2)$
on the integration variable $M^2$ requires numerical integration of the 
correction terms (\ref{K1}) to (\ref{K4}).
\footnote{A computer program for the evaluation of the photoabsorption cross
section according to (A.13), (A.14) and (A.19) to (A.22) is available.}
The correction terms (\ref{K1}) and (\ref{K2}) will be analysed in
Appendix C.

A considerable simplification of the various contributions to the cross
sections in (\ref{A15}) and (\ref{A16}), as well as in (\ref{K3})
and (\ref{K4}), is obtained by a restriction to realistic values of
the parameters $m^2_0, m^2_1 (W^2) = \xi \Lambda^2_{sat}(W^2)$ and
$\Lambda^2_{sat} (W^2)$. Expansion in terms of 
\be
\mu (W^2) = \frac{m^2_0}{\Lambda^2_{sat} (W^2)} \ll 1
\label{A25}
\ee
and
\be
\frac{1}{\xi} \ll 1
\label{A26}
\ee
will then lead to the photoabsorption cross sections in (\ref{2.33a}) of
the main text.

We proceed in two steps. In a first step, we analyse the cross sections
in the color-transparency limit of $\eta (W^2,Q^2) \gg 1$. Taking into
account the leading terms in the expansion in $\mu (W^2), 1/\xi$ and
$1/\eta (W^2,Q^2)$, the cross sections for the dominant terms in (\ref{A15})
and (\ref{A16}) take the simple form
\bqa
\sigma^{dom}_{\gamma^*_L p} (W^2,Q^2) & = & A(W^2) \frac{1}{6 \eta}
\frac{u^3 + 3u^2 + 6u}{(1+u)^3} 
\label{A27}
\eqa
and
\bqa
\sigma^{dom}_{\gamma^*_T p} (W^2,Q^2) & = & A(W^2) \frac{1}{3 \eta}
\frac{2u^3 + 6u^2}{2(1+u)^3},
\label{A28}
\eqa
where $u$ denotes the ratio
\be
u \equiv u (\eta (W^2,Q^2)) = \frac{\xi}{\eta (W^2,Q^2)} =
\frac{m^2_1 (W^2)}{Q^2 + m^2_0}.
\label{A29}
\ee

The $m^2_1 (W^2)$ correction terms, given by the integrals (\ref{K3})
and (\ref{K4}), are treated as follows. We replace the integration variable
$M^2$ by $0 \le x^\prime \le 1$, according to
\bqa
M^2 &=& (2m_1 \Lambda - \Lambda^2) x^\prime + (m_1 - \Lambda)^2 \nonumber \\
& = & \xi \Lambda^2 \left(1 - \frac{2}{\sqrt{\xi}} (1-x^\prime) +
\frac{1}{\xi} (1-x^\prime) \right),
\label{A30}
\eqa
and expand the integrands in powers of the small parameters $1/\xi,~
\mu (W^2)$ and $1/\eta (W^2,Q^2)$. A somewhat lengthy analysis, upon 
carrying out the integration over $dx^\prime$, yields the $m^2_1 (W^2)$
correction terms, 
\be
\Delta \sigma_{\gamma^*_L p}^{(m^2_1)} (W^2,Q^2)  =  - A(W^2) \frac{1}{6 \eta}
\frac{6u}{(1+u)^3}, 
\label{A31}
\ee
and
\be
\Delta \sigma_{\gamma^*_T p}^{(m^2_1)} (W^2,Q^2) =  - A(W^2) \frac{1}{3 \eta}
\frac{3u^2 - 3u}{2(1+u)^3}.
\label{A32}
\ee
Addition of (\ref{A27}) and (\ref{A31}), as well as addition of
(\ref{A28}) and (\ref{A32}), yields
\be
\sigma_{\gamma^*_L p} (W^2,Q^2) =  A(W^2) \frac{1}{6 \eta} G_L (u),
\label{A33}
\ee
and
\be
\sigma_{\gamma^*_T p} (W^2,Q^2) =  A(W^2) \frac{1}{3 \eta} G_T (u),
\label{A34}
\ee
where \cite{PRD85} 
\be
G_L (u) =  \frac{2u^3 + 6u^2}{2(1+u)^3}, 
\label{A35}
\ee
and
\be
G_T (u)  =  \frac{2u^3 + 3u^2 + 3u}{2(1+u)^3}.
\label{A36}
\ee
compare (\ref{2.34}) in the main text.

So far we analysed the cross sections (\ref{A3}) and (\ref{A4}) for 
$\eta (W^2,Q^2) \gg 1$.
The dependence of the cross sections on the upper limit $m^2_1 (W^2)$ for
$\eta (W^2,Q^2)> 1$ turned out to be given by the correction factors 
(\ref{A35}) and (\ref{A36}) that
depend  on the ratio \break
$u(\eta (W^2,Q^2)) = \xi/\eta (W^2,Q^2)$ in (\ref{A29}). With the assumption 
that
the dependence on $m^2_1 (W^2)$ also for $\eta (W^2,Q^2) < 1$ is 
determined by a correction factor that is an 
analytic function 
of this ratio 
$u (\eta (W^2,Q^2)) = \xi/\eta (W^2,Q^2)$, the correction factors in the
extension of the cross sections (\ref{A33}) and (\ref{A34}) to 
values of $\eta (W^2,Q^2) < 1$ have to coincide with 
$G_L (u)$ and $G_T (u)$ in (\ref{A35}) and (\ref{A36}). In (\ref{A33})
and (\ref{A34}), 
accordingly, we have to perform the
replacements
\bqa
\frac{1}{6 \eta} & \to & I^{(1)}_L (\eta, \mu), \nonumber \\
\frac{1}{3 \eta} & \to & I^{(1)}_T (\eta, \mu),
\label{A37}
\eqa
leading to 
\bqa
\sigma_{\gamma^*_L p} (W^2,Q^2) & = & A (W^2) I^{(1)}_L (\eta, \mu) G_L (u),
\nonumber \\
\sigma_{\gamma^*_T p} (W^2,Q^2) & = & A (W^2) I^{(1)}_T (\eta, \mu) G_T (u),
\label{A38}
\eqa
where the general $m^2_1 \to \infty$ expression for 
$I^{(1)}_L (\eta, \mu)$ and $I^{(1)}_T (\eta, \mu)$ are given by 
(\ref{2.9}) or (\ref{2.14}) in the main text. The result (\ref{A38}), 
upon inserting the definition (\ref{A17}) for $A(W^2)$, 
yields (\ref{2.33a}) 
(for $\rho = 1$) in the main text.

In fig. A1, we illustrate the dependence of $\sigma_{\gamma^*p}
(W^2,Q^2) = \sigma_{\gamma^*p} (\eta (W^2,Q^2), \xi)$ on the parameter
$\xi$ introduced in (\ref{2.30}) via $M^2_{q \bar q} \le m^2_1 (W^2) =
\xi \Lambda^2_{sat} (W^2).$ For $\eta (W^2,Q^2) \lsim 1$, high-mass 
fluctuations of the photon, $\gamma^* \to q \bar q$, do not contribute, and
the limit of $\xi \to \infty$ becomes valid.

\begin{figure}[t]
\begin{center}
\epsfig{file=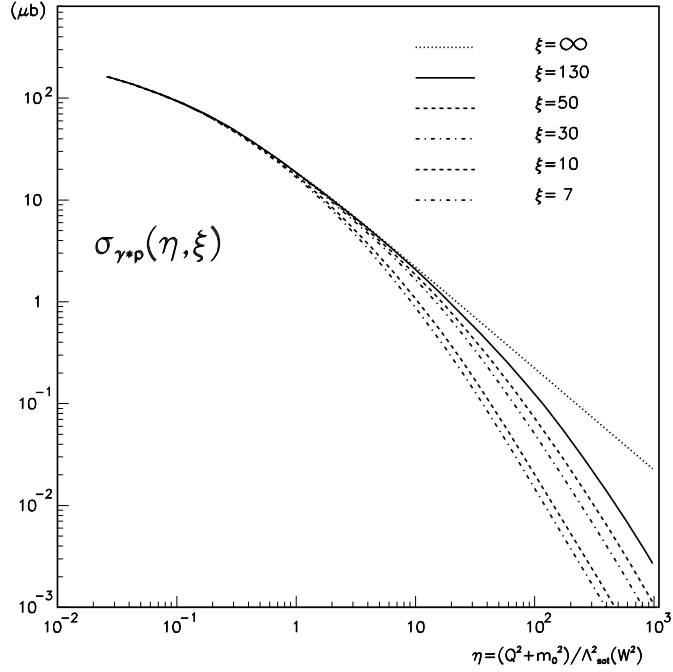,height=10cm, width=10cm}
\caption{The effect on the total photoabsorption cross section of the
exclusion of high-mass $q \bar q$ fluctuations via $M^2_{q \bar q} \le
\xi \Lambda^2_{sat} (W^2)$.}
\end{center}
\end{figure}

We turn to the transverse-size-enhancement factor $\rho$ introduced in
(\ref{2.24}) and (\ref{2.33}). Quark-antiquark $(q \bar q)$ states of
given mass $M_{q \bar q}$ originating from longitudinally and 
transversely polarized photons, $\gamma^*_L \to (q \bar q)_L$ and
$\gamma^*_T \to (q \bar q)_T$, differ in their internal transverse
momentum distributions. The normalized $z(1-z)$ distributions 
\cite{Ku-Schi}, \cite{PRD85}
\be
f_L (z(1-z)) = 6 z(1-z)
\label{A39}
\ee
and
\be
f_T (z(1-z)) = \frac{3}{2} (1-2z (1-z))
\label{A40}
\ee
imply different average transverse momenta of the quarks originating from
$\gamma^*_L \to (q \bar q)_L$ and $\gamma^*_T \to (q \bar q)_T$ transitions.
Passing from the ratio of the average
internal transverse momenta squared of the quarks (antiquarks) in the
$(q \bar q)^{J=1}_L$ and $(q \bar q)^{J=1}_T$ states to the ratio of the transverse
sizes of the $(q \bar q)^{J=1}_L$ and $(q \bar q)^{J=1}_T$ states, 
from (\ref{A39}) and (\ref{A40}), by invoking the
uncertainty principle, one finds an enhancement in the average transverse size of
magnitude \cite{Ku-Schi}, \cite{PRD85}
\be
\rho = \frac{4}{3}
\label{A41}
\ee
of the $(q \bar q)^{J=1}_T$ relative to the $(q \bar q)^{J=1}_L$ state.
The average  transverse size of $q \bar q$ fluctuations being unaffected by the
Lorentz-boost transformation supplying the $\gamma^*p$ interaction 
energy $W$, the ratio $\rho$ quite generally, i.e. independently of the
specific value of $\rho = 4/3$ in (\ref{A41}), must be a W-independent
constant,
$\rho = const.$

The color-gauge-invariant interaction (\ref{2.2}) in terms of
$(q \bar q)^{J=1}_{L,T}$ states, in the appropriate high-energy limit
of $\vec l^{~\prime 2}_{\bot~Max}~\vec r^{~\prime 2}_\bot \ll 1$ implies
vanishing of the dipole cross section in the limit of vanishing dipole
size, or proportionality to $\vec r^{~\prime 2}_\bot$ \cite{PRD85},
\be
\bar \sigma_{(q \bar q)^{J=1}_{L,T} p} (\vec r^{~\prime 2}_\bot, W^2) =
\frac{1}{4} \pi \vec r^{~\prime 2}_\bot \int d \vec l^{~\prime 2}_\bot
\vec l^{~\prime 2}_\bot \bar \sigma_{(q \bar q)^{J=1}_{L,T} p} 
(\vec l^{~\prime 2}_\bot, W^2).
\label{A42}
\ee
The factor multiplying $\vec r^{~\prime 2}_\bot$ on the right-hand side
in (\ref{A42}), for the dipole cross section of the ansatz (\ref{2.3}),
or equivalently according to the ansatz (\ref{A14}), is proportional to
$\Lambda^2_{sat} (W^2)$. A transverse-size enhancement of magnitude
$\rho$, corresponding to an enhancement of the dipole cross section for
$(q \bar q)^{J=1}_T$ states, accordingly implies the replacement
\be
\Lambda^2_{sat} (W^2) \to \rho \Lambda^2_{sat} (W^2)
\label{A43}
\ee
to be carried out in (\ref{A14}), modifying
(\ref{A14}) to become
\bqa
\bar \sigma_{(q \bar q)^{J=1}_L p} (\vec l^{~\prime 2}_\bot, W^2) & = &
\frac{\sigma^{(\infty)} (W^2)}{\pi} \delta (\vec l^{~\prime 2}_\bot -
\Lambda^2_{sat} (W^2)), \nonumber \\
\bar \sigma_{(q \bar q)^{J=1}_T p} (\vec l^{~\prime 2}_\bot, W^2) & = &
\frac{\sigma^{(\infty)} (W^2)}{\pi} \delta (\vec l^{~\prime 2}_\bot - \rho
\Lambda^2_{sat} (W^2)).
\label{A44}
\eqa
Inserting (\ref{A44}) into (\ref{A42}) yields (\ref{2.45}) and
(\ref{2.44}) in the
main text.

Returning to the photoabsorption cross sections in (\ref{A38}), we
apply the substitution (\ref{A43}) to the transverse cross section via
\be
I^{(1)}_T (\eta, \mu) \to I^{(1)}_T \left( \frac{\eta}{\rho}, 
\frac{\mu}{\rho} \right)
\label{A45}
\ee
yielding (\ref{2.33}) and (\ref{2.33a}) in the main text.\footnote{One
may alternatively be led to the conclusion that also $G_T (u) \to
G_T (u = \xi/\eta \to \rho \xi/\eta)$ in (\ref{A36}). Effectively this
would amount to a different value of the parameter $\xi$ in the
transverse case relative to the longitudinal one. The replacement (\ref{A45})
is identical to the approach of ref. \cite{PRD85}, see next paragraph.}

The introduction of the transverse-size-enhancement factor $\rho$
according to (\ref{A39}) to (\ref{A41}), leading to (\ref{A44}), 
makes use of the dipole
cross sections for longitudinally and transversely polarized dipole states,
$(q \bar q)^{J=1}_L$ and $(q \bar q)^{J=1}_T$ in (\ref{A14}). In ref. \cite{PRD85},
we introduced an ansatz for the dipole cross section $\sigma_{(q \bar q)}
(\vec r_\bot, z(1-z), W^2)$ in (\ref{2.1}) that replaces the ansatz
(\ref{A14}), and equivalently
(\ref{2.4}), and contains the cases of $\rho = 1$, as well as $\rho \not= 1$,
and specifically $\rho = 4/3$ from (\ref{A41}).
The factor $\rho$ depends on a free parameter, $\rho = \rho (\epsilon \equiv
1/6a)$. Explicitly,
\be
\rho (\epsilon) = \frac{1}{2 \sqrt{1-4 \epsilon}} \left( \ln
\frac{(1 + \sqrt{1-4 \epsilon})^2}{4 \epsilon} - \sqrt{1 - 4 \epsilon} \right)
\cong \frac{1}{2} \ln \frac{1}{\epsilon}.
\label{A46}
\ee
The photoabsorption cross sections (\ref{2.33a}) to (\ref{2.35}) of the
present paper form an accurate closed approximate form of the results in
ref.\,\cite{PRD85}. The results of ref.\,\cite{PRD85} explicitly demonstrate
the consistency of the photoabsorption cross sections in (\ref{2.33a}) to
(\ref{2.35}) with the general form of the CDP as formulated in (\ref{2.1})
with (\ref{2.2}).
\end{appendix}

\begin{appendix}

\renewcommand{\theequation}{\Alph{section}.\arabic{equation}}
\setcounter{section}{2}
\setcounter{equation}{0}
\section*{Appendix B. The Mass Dispersion Relation of Generalized Vector
Dominance}
In this Appendix, we explicitly demonstrate the connection between
the CDP and the mass dispersion relation of the generalized vector dominance
(GVD) approach \cite{Sakurai} of the 1970's.We consider the total 
photoabsorption cross section given by the sum of the longitudinal
and transverse cross sections in (\ref{A3}) and (\ref{A4}). We restrict
ourselves to the
case of helicity independence of the $q \bar q$-proton interaction,
\be
\bar \sigma_{(q \bar q)^{J=1}_L p} (\vec l^{~\prime 2}_\bot, W^2) =
\bar \sigma_{(q \bar q)^{J=1}_T p} (\vec l^{~\prime 2}_\bot, W^2) \equiv
\bar \sigma_{(q \bar q)^{J=1} p} (\vec l^{~\prime 2}_\bot, W^2)
\label{B1}
\ee
that is realized by the ansatz (\ref{2.3}). Summation of
(\ref{A3}) and (\ref{A4}) 
%\bqa
%&& \frac{M^2}{(Q^2+M^2)^2} + \frac{M^{\prime 2}}{(Q^2+M^{\prime 2})^2} -
%\frac{M^2 + M^{\prime 2} - \vec l^{~\prime 2}_\bot}{(Q^2+M^2)
%(Q^2+M^{\prime 2})} = \nonumber \\
%&& = \frac{-Q^2}{(Q^2+M^2)^2} - \frac{Q^2}{(Q^2+M^{\prime 2})^2} +
%\frac{2Q^2}{(Q^2+M^2)(Q^2+M^{\prime 2})},
%\label{B2}
%\eqa
yields
\bqa
&& \hspace*{-1cm}\sigma_{\gamma^*p} (W^2,Q^2) = \sigma_{\gamma^*_L p} 
(W^2,Q^2) +\sigma_{\gamma^*_T p} (W^2,Q^2) =  \frac{\alpha R_{e^+e^-}}{6 \pi}  
     \nonumber \\
&& \hspace*{-1cm} \times \int dM^2 \int dM^{\prime 2}
\int d \vec l^{~\prime 2}_\bot \bar \sigma_{(q \bar q)^{J=1} p} 
(\vec l^{~\prime 2}_\bot , W^2) \frac{w(M^2,M^{\prime 2},
\vec l^{~\prime 2}_\bot) (M^{\prime 2} - M^2 +\vec l^{~\prime 2}_\bot)}
{(Q^2+M^2)(Q^2+M^{\prime 2})}.
\label{B3}
\eqa
Substituting the $J=1$ projections of the ansatz (\ref{2.3}),
\bqa
\bar \sigma_{(q \bar q)^{J=1}_L p} (\vec l^{~\prime 2}_\bot, W^2) & = &
\bar \sigma_{(q \bar q)^{J=1}_T p} (\vec l^{~\prime 2}_\bot, W^2) =
\nonumber \\
& = & \frac{\sigma^{(\infty)} (W^2)}{\pi} \delta \left( \vec l^{~\prime 2}_\bot
- \Lambda^2_{sat} (W^2) \right),
\label{B4}
\eqa
(\ref{B3}) becomes
\bqa
\hspace*{-0.8cm}
\sigma_{\gamma^*p} (W^2,Q^2) & = & \frac{\alpha R_{e^+e^-}}{6 \pi^2}
\sigma^{(\infty)} (W^2) \nonumber \\
& \times & \int dM^2 \int dM^{\prime 2}
\frac{w \left(M^2, M^{\prime 2}, \Lambda^2_{sat} (W^2) \right) 
(M^{\prime 2} - M^2 + \Lambda^2_{sat} 
(W^2))}{(Q^2 + M^2) (Q^2+M^{\prime 2})}.
\label{B5}
\eqa
The right-hand sides in (\ref{B3}) and (\ref{B5}) are identical in
their structures to the mass-dispersion relation of the GVD approach of the
1970's \cite{Sakurai}, \cite{FR}. In the 1970's, the mass dispersion relation
was postulated by extrapolating the role of the low-lying vector mesons
$\rho^0, \omega$ and $\phi$
in $e^+e^-$ annihilation and (diffractive)
photoproduction to a conjectured continuum
of high-mass-vector-state contributions predicted to be observed in
$e^+e^-$ annihilation experiments at sufficiently high energies. The approach
in ref. \cite{Sakurai} is based on the simplifying assumption of the
``diagonal approximation'', $M^2 = M^{\prime 2}$ in (\ref{B5}). The
destructive interference of the QCD gauge theory contained in (\ref{2.2}),
in ref. \cite{FR}
was anticipated by introducing off-diagonal,
$M^2 \not= M^{\prime 2}$, transitions, compare (\ref{B5}).
\end{appendix}

\begin{appendix}

\renewcommand{\theequation}{\Alph{section}.\arabic{equation}}
\setcounter{section}{3}
\setcounter{equation}{0}
\renewcommand{\thetable}{\Alph{section}.\arabic{table}}
\setcounter{section}{3}
\setcounter{table}{0}
\section*{Appendix C. The \boldmath$m^2_0$\unboldmath ~
correction in (\ref{2.8}) and (\ref{A11}).}
We  discuss the (relative) magnitude of the $m^2_0$-dependent
corrections,\break $\Delta \sigma^{(m^2_0)}_{\gamma^*_{L,T}p}  (W^2,Q^2)$, to
the total photoabsorption cross section, $\sigma_{\gamma^*_{L,T}p} (W^2,Q^2)$
in (\ref{2.7}) with (\ref{2.8}),
as introduced according to (\ref{A11}) and specified in (\ref{K1}) and 
(\ref{K2}). From the restricted range of the integrations in (\ref{K1})
and (\ref{K2}), we expect contributions of order $\mu (W^2) = 
m^2_0/\Lambda^2_{sat} (W^2) \ll 1$.

Rewriting $\Delta \sigma^{(m_0^2)}_{\gamma^*_{L,T}p} (W^2,Q^2)$ 
from (\ref{K1}) and (\ref{K2}) as
\be
\Delta \sigma^{(m_0^2)}_{\gamma^*_{L,T}p} (W^2,Q^2) = A(W^2) 
I^{(1)}_{L,T} (\eta, \mu) \delta^{(m^2_0)}_{L,T} (\eta, \mu),
\label{C1}
\ee
with $\eta \equiv \eta(W^2,Q^2)$ and $\mu \equiv \mu (W^2)$, and with
$A(W^2)$ from (\ref{A17}) and $I^{(1)}_{L,T} (\eta, \mu)$ from
(\ref{2.9}), and introducing the $Q^2 = 0$ photoproduction limit,
$\sigma_{\gamma p} (W^2)$, according to 
(\ref{2.40}), the total photoabsorption cross section becomes
\bqa
\sigma_{\gamma^*p} (W^2,Q^2) & = &\frac{\sigma_{\gamma p} (W^2)}{I^{(1)}_T
(\mu, \mu) (1+\delta^{(m^2_0)}_T (\mu, \mu))}  \nonumber \\
& \times &\left(
I^{(1)}_T (\eta, \mu) (1+\delta^{(m^2_0)}_T (\eta, \mu)) +
I^{(1)}_L (\eta, \mu) (1+\delta^{(m^2_0)}_L (\eta, \mu)) \right).
\label{C2}
\eqa
It may be rewritten as
\bqa
\hspace*{-0.8cm}
\sigma_{\gamma^*p} (W^2,Q^2) & = & \frac{\sigma_{\gamma p} (W^2)}{\ln
  \frac{1}{\mu}} \left(I^{(1)}_T (\eta, \mu) + I^{(1)}_L (\eta, \mu)\right)
\nonumber \\
& \times & \frac{1}{1 + \delta^{(m^2_0)}_T (\mu, \mu)}
\left( 1+ \frac{I^{(1)}_T (\eta, \mu) \delta^{(m^2_0)}_T (\eta, \mu) +
I^{(1)}_L (\eta, \mu) \delta^{(m^2_0)}_L (\eta, \mu)}{I^{(1)}_T (\eta, \mu)
+ I^{(1)}_L (\eta, \mu)} \right).
\label{C3}
\eqa
The (dominant) contribution to $\sigma_{\gamma^*p} (W^2,Q^2)$, on the
right-hand side of the first line in (\ref{C3}), is corrected by the 
$m^2_0$-correction term shown in the second line of (\ref{C3}).

In the transition from (\ref{C2}) to (\ref{C3}), we used $I_T^{(1)}
(\mu, \mu) = \ln (1/\mu)$ from (\ref{2.10}). 
Upon specification of
(\ref{K2}) to $Q^2 = 0$, one finds that $\delta_T^{(m^2_0)} (\mu, \mu)$
in (\ref{C1}) is approximately given by
\be
\delta^{(m^2_0)}_T (\mu, \mu) \cong - \frac{\mu}{\ln \frac{1}{\mu}} < 0.
\label{C4}
\ee
In Table C.1, we compare the approximation (\ref{C4}) with the 
numerical evaluation of 
$\delta_T^{(m^2_0)} (\mu, \mu)$ for several values of $W^2$ and
$m^2_0 = 0.15 GeV^2$. The error of the approximation (\ref{C4}) decreases
from about 9 \% to about 1 \% in the energy range considered in Table C.1.
\begin{table}
\begin{center}
\begin{tabular}{|r|r|r|}
\hline
$W^2$(GeV$^2$) & $\delta_T^{(m_0^2)}(\mu,\mu)$ & $-{\mu\over{\ln{1\over\mu}}}$ \\
\hline
  50  & $-1.03\times 10^{-1}$  & $-9.44\times 10^{-2}$ \\
 100  & $-7.60\times 10^{-2}$  & $-7.09\times 10^{-2}$ \\
 625  & $-3.59\times 10^{-2}$  & $-3.45\times 10^{-2}$ \\
 $1\times 10^4$ & $-1.27\times 10^{-2}$ & $-1.25\times 10^{-2}$ \\
 $4\times 10^4$ & $-7.79\times 10^{-3}$ & $-7.72\times 10^{-3}$ \\
\hline
\end{tabular}
\caption{Comparison of the numerical results for $\delta^{(m^2_0)}_T 
(\mu, \mu)$ from (\ref{K2}) and (\ref{C1}) with the analytical 
approximation (\ref{C4}).}
\end{center}
\vspace*{-0.5cm}
\end{table}

The $m^2_0$ correction term in the bracket on the right-hand side in
(\ref{C3}),
\be
\delta^{(m^2_0)} (\eta, \mu) \equiv \frac{I^{(1)}_T (\eta, \mu) 
\delta_T^{(m^2_0)} (\eta, \mu) + I^{(1)}_L (\eta, \mu) \delta^{(m^2_0)}_L
(\eta, \mu)}{I^{(1)}_T (\eta, \mu) + I^{(1)}_L (\eta, \mu)},
\label{C5}
\ee
as well as $\delta^{(m^2_0)}_T (\mu, \mu)$, must be
evaluated by numerical integrations in (\ref{K1})
and (\ref{K2}).
In Table C.2, we present the numerical results for $\sigma_{\gamma^*p}
(W^2,Q^2)$ in (\ref{C3}) by discriminating between the dominant contribution
based on evaluating (\ref{2.9}) with (\ref{2.10}), and the correction term
in the second line in
(\ref{C3}). To be definite, for $\sigma_{\gamma p} (W^2)$ 
in (\ref{C3}), for the results in Table C.2 we used
the PDG fit from (\ref{2.41}), and the parameters
specified in (\ref{3.5}) and (\ref{3.6}),
\begin{eqnarray}
C_1 & = & 0.31,\nonumber \\
C_2 & = & 0.27,\nonumber\\
m^2_0 & = & 0.15 GeV^2,\nonumber \\
\xi & = & 130.
\label{C6}
\end{eqnarray}

\begin{table}
\begin{center}
\begin{tabular}{|r|r|l|r|l|}
\hline
$\eta(W^2,Q^2)$ & $Q^2$(GeV$^2$) & $\sigma_{\gamma^*p}^{(dom)}(W^2,Q^2)$ ($\mu$b)&
   ${{1+\delta^{(m_0^2)}(\eta,\mu)}\over{1+\delta_T^{(m_0^2)}(\mu,\mu)}}$ &
   $\sigma_{\gamma^*p}(W^2,Q^2)$ ($\mu$b)\\
  $W=225$GeV &   &    &   &  \\
\hline
  0.026 &   0  & $1.63\times 10^2$ & 1.000 
        & $1.63\times 10^2$   \\
  0.1   & $4.28\times 10^{-1}$ & $9.47\times 10^1$ & 1.006  
        & $9.53\times 10^1$ \\ 
     1  & $5.63\times 10^0$ & $1.90\times 10^1$ & 1.007  
        & $1.92\times 10^1$ \\
  10    & $5.76\times 10^1$ &  2.19 & 1.007  
        & 2.20 \\
  100   & $5.77\times 10^2$ & $2.23\times 10^{-1}$ & 1.007
        & $2.24\times 10^{-1}$ \\
  1000  & $5.63\times 10^3$ & $2.29\times 10^{-2}$ & 1.007
        & $2.31\times 10^{-2}$ \\
\hline
\hline
\hline
$\eta(W^2,Q^2)$ & $Q^2$(GeV$^2$) & $\sigma_{\gamma^*p}^{(dom)}(W^2,Q^2)$ ($\mu$b)&
   ${{1+\delta^{(m_0^2)}(\eta,\mu)}\over{1+\delta_T^{(m_0^2)}(\mu,\mu)}}$ &
   $\sigma_{\gamma^*p}(W^2,Q^2)$ ($\mu$b)\\
  $W=25$GeV &   &    &   &  \\
\hline
  0.085 &   0  & $1.19\times 10^2$ & 1.000 
        & $1.19\times 10^2$   \\
  0.1   & $2.63\times 10^{-2}$ & $1.08\times 10^2$ & 1.011  
        & $1.09\times 10^2$ \\ 
     1  & $1.61\times 10^0$ & $2.02\times 10^1$ & 1.035  
        & $2.10\times 10^1$ \\
  10    & $1.75\times 10^1$ &  2.36 & 1.038  
        & 2.45 \\
\hline
\end{tabular}
\caption{The results for the total photoabsorption cross section including
the $m^2_0$ correction, showing that the $m^2_0$ correction
in (\ref{C3}) and (\ref{C5}) can safely
be neglected in the numerical fits to the experimental data for the total
photoabsorption cross section.}
\end{center}
\end{table}

\begin{table}
\begin{center}
\begin{tabular}{|r | r|r|}
\hline
 $Q^2$(GeV$^2$) &  $R^{(dom)}(W^2,Q^2)$ & $R(W^2,Q^2)$ \\
\hline
  0.001 & 0.001 &  0.001 \\
  0.01  & 0.013 &  0.014 \\ 
  0.1   & 0.101 &  0.108 \\
  1.0   & 0.319 &  0.343 \\
  10    & 0.457 &  0.487 \\
  100   & 0.494 &  0.501 \\
  1000  & 0.499 &  0.500 \\
\hline
\end{tabular}
\caption{The ratio $R(W^2,Q^2)$ from (\ref{C7}) for $W^2 = (200)^2~ GeV^2$.}
\end{center}
\end{table}

The fairly small numerical values of the correction factor
in Table C.2 are partially due to a cancellation between the contributions
of $\delta_T^{(m^2_0)} (\eta, \mu)$ and $\delta_L^{(m^2_0)} (\eta, \mu)$ 
in (\ref{C5}). The
corrections $\delta_T^{(m^2_0)} (\eta, \mu)$ 
and $\delta_L^{(m^2_0)} (\eta, \mu)$ are of opposite
sign. The effect on the ratio
\bqa
R(W^2,Q^2) & = & \frac{\sigma_{\gamma^*_Lp} (W^2,Q^2)}{\sigma_{\gamma^*_Tp}
(W^2,Q^2)} = \frac{I^{(1)}_L (\eta, \mu) (1 + \delta_L^{(m^2_0)} (\eta, \mu))}
{I^{(1)}_T (\eta, \mu) (1 + \delta_T^{(m^2_0)} (\eta, \mu))} \nonumber
\\
& \equiv & R^{(dom)} (W^2,Q^2) \frac{1 + \delta_L^{(m^2_0)}(\eta, \mu)}
{1 + \delta_T^{(m^2_0)}(\eta, \mu)}
\label{C7}
\eqa
is more significant, accordingly. 

In Table C.3, we show a few values of $R (W^2,Q^2)$ without the 
$m^2_0$-correction term, $R^{(dom)} (W^2,Q^2)$, and with the $m^2_0$
correction according to (\ref{C7}). Table C.3 shows the expected convergence
to $R(W^2,Q^2) = 1/2$ according to (\ref{2.23}).

In summary, the analysis of this Appendix shows that the $m^2_0$-correction
terms appearing in (\ref{A11}), and indicated as additional terms of
$O (\mu)$ in (\ref{2.8}), may be neglected in the fit to the 
experimental data.

\end{appendix}

%\centerline{\Large\bf Acknowledgement}

\end{document}